\documentclass[aps,prb,showpacs,preprintnumbers,amsmath,amssymb,twocolumn,floatfix,footinbib,superscriptaddress,longbibliography]{revtex4-2}

\usepackage{amsmath,amssymb}
\usepackage{graphicx}
\usepackage{bm}
\usepackage{bbm}
\usepackage{color}
\usepackage{natbib}
\usepackage{hyperref}
\hypersetup{
  colorlinks,
  citecolor=magenta,
  linkcolor=blue,
  urlcolor=blue}

\newcommand{\gbf}[1]{\boldsymbol #1}
\newcommand{\I}{\mathrm{i}}
\newcommand{\E}{\mathrm{e}}
\newcommand{\dd}{\mathrm{d}}
\newcommand{\nn}{\nonumber}
\newcommand{\cfig}[1]{Fig.~\ref{#1}}
\newcommand{\csec}[1]{Sec.~\ref{#1}}
\newcommand{\ceqn}[1]{Eq.~(\ref{#1})}
\newcommand{\capp}[1]{App.~\ref{#1}}

\newcommand{\cref}[1]{Ref.~\onlinecite{#1}}

\newcommand{\pd}{\phantom{\dagger}}

\begin{document}
\title{Chiral Heisenberg Gross-Neveu-Yukawa criticality:\\ Honeycomb vs. SLAC fermions}

\author{Thomas C. Lang}
\email{thomas.lang@uibk.ac.at}
\affiliation{Institute for Theoretical Physics, University of Innsbruck, 6020 Innsbruck, Austria}
\author{Andreas M. L\"{a}uchli}
\affiliation{Laboratory for Theoretical and Computational Physics,
Paul Scherrer Institute, 5232 Villigen, Switzerland}
\affiliation{Institute of Physics, \'{E}cole Polytechnique F\'{e}ed\'{e}rale de Lausanne (EPFL), 1015 Lausanne, Switzerland}

\begin{abstract}
We perform large scale quantum Monte Carlo simulations of the Hubbard model at half filling with a single Dirac cone close to the critical point, which separates a Dirac semi-metal from an antiferromagnetically ordered phase where SU(2) spin rotational symmetry is spontaneously broken. We discuss the implementation of a single Dirac cone in the SLAC formulation for eight Dirac components and the influence of dynamically induced long-range super-exchange interactions. The finite-size behavior of dimensionless ratios and the finite-size scaling properties of the Hubbard model with a single Dirac cone are shown to be superior compared to the honeycomb lattice. We extract the critical exponent believed to belong to the chiral Heisenberg Gross-Neveu-Yukawa universality class: The critical exponent ${\nu = 1.02(3)}$ coincides for the two lattice types once honeycomb lattices of linear dimension ${L>15}$ are considered. In contrast to the SLAC formulation, where the anomalous dimensions are estimated to be ${\eta_{\phi}=0.73(1)}$ and ${\eta_{\psi}=0.09(1)}$, they remain less stable on honeycomb lattices, but tend towards the estimates from the SLAC formulation. 
\end{abstract}

\maketitle

\section{Introduction}
%
The family of Gross-Neveu-Yukawa (GNY) universality classes capture the complex interplay of critical Dirac fermions coupled to a scalar bosonic fields. It represents the simplest universality of quantum critical points that lack classical analogues and realize the most fundamental scenario of relativistic fermions: the spontaneous acquisition of mass triggered by interactions \cite{Gross74,ZinnJustin91}. Here we focus specifically on the chiral Heisenberg [O(3) or SU(2)] transition \cite{Herbut09a} with ${N=8}$ Dirac components in $(2\!+\!1)$ dimensions found, e.g., in semi-metal to insulator transitions, which remains hard to control in analytical approaches \cite{Janssen14,Zerf17,Gracey18,Knorr18,Ladovrechis23,TolosaSimeon25} as well as quantum Monte Carlo (QMC) simulations \cite{Assaad13,Toldin15,Otsuka16,Tang18,Buividovich18,Buividovich19,Liu19,Otsuka20,Ostmeyer20,Liu21,Ostmeyer21b,Xu21,Otsuka22,Yu23}. Where the critical properties of the chiral Ising GNY transition have been arguably consistently determined to be at archipelago provided by conformal bootstrap \cite{Erramilli23}, competing interaction channels and generically large corrections in finite-size scaling prevent the same for the chiral Heisenberg fixed point \cite{Ladovrechis23}. 

\begin{figure}[tp]
  \centering
  \includegraphics[width=\columnwidth]{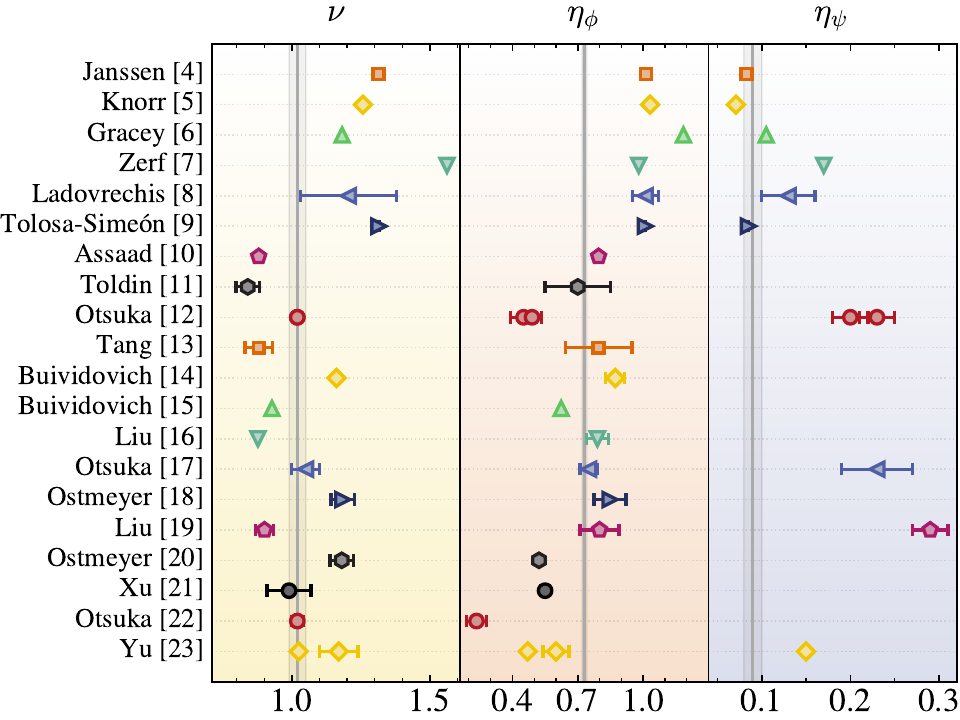}\\
  \caption{Comparison of estimates for the correlation-length exponent $\nu$, the boson anomalous dimension $\eta_{\phi}$ and the fermion anomalous dimension $\eta_{\psi}$, assuming the dynamic scaling exponent ${z=1}$, from analytical approaches \cite{Janssen14,Zerf17,Gracey18,Knorr18,Ladovrechis23,TolosaSimeon25} and Monte Carlo simulations \cite{Assaad13,Toldin15,Otsuka16,Tang18,Buividovich18,Buividovich19,Liu19,Otsuka20,Ostmeyer20,Liu21,Ostmeyer21b,Xu21,Otsuka22,Yu23} for the ${N=8}$ chiral-Heisenberg universality class in chronological order. Gray vertical bars indicate the estimates and their errorbars from the SLAC-Hubbard model discussed in this manuscript.}
  \label{comparison}
\end{figure}

The fairly disparate state of estimates for the ${N=8}$ critical exponents which allow us to classify the universality of the chiral Heisenberg transition is summarized in \cfig{comparison} along with the estimates from this investigation. In particular the results from QMC simulations, which rightfully pride themselves to be numerically exact, are far from a consensus. This illustrates the ambiguities in the finite-size scaling analysis and the limitations of practically accessible spacetime volumes.

Motivated by simulations in the Hamiltonian formulation of SLAC fermions \cite{Drell76,Li18,Lang19,Tabatabaei22} in this manuscript we present the implementation and benchmark simulation results for the chiral-Heisenberg GNY quantum phase transition of the fermionic SLAC Hubbard model with $N=8$ Dirac components, which allows for a direct comparison with simulations on conventional lattices such as the honeycomb, or the $\pi$-flux lattice. We argue that the spread of the QMC results can be attributed to strong finite-size corrections in conventional lattice implementations and show that our simulations, which maximize the linear Dirac dispersion across the Brillouin zone, significantly reduce finite-size effects at the critical point. This allows for computationally favorable and more accurate estimates of the physics at criticality. 

In the following \csec{Implementation} we describe the implementation and simulation details of congruent Dirac cones with a particular number $N$ of zero modes on the lattice and address the reliability and potential issues such as locality, renormalizability and dynamically induced non-local interactions in \csec{locality}. A finite-size scaling analysis and estimates for the critical exponents are presented in \csec{FSS}. In the following we indicate results obtained with the SLAC formulation with a {\includegraphics[height=\fontcharht\font`\B]{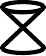}}$\,$-icon, where the inscribed number denotes the number of massless modes $N$. Results obtained on the honeycomb lattice (${N=8}$) are identified with a {\includegraphics[height=\fontcharht\font`\B]{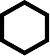}}$\,$-icon.

\section{Model \& method\label{Implementation}}

In this section we explain the practical implementation and properties of a generalization of SLAC fermions \cite{Drell76}, in particular for the comparison with electrons on the honeycomb lattice. 

\subsection{Formulation for \textit{N}-pole SLAC fermions}

A Dirac cone carries $N$ poles of the massless momentum space Dirac propagator, which results in $N$ zero modes on the lattice. In two spatial dimensions the Dirac operator with Fermi velocity $v_\text{F}$ may be represented by the ${2\times 2}$ Hamiltonian
\begin{align}
   H_{\mathbf{k}\sigma} = - v_\text{F} \sum_{\mathbf{k}}\gbf{\psi}^{\dagger}_{\mathbf{k}\sigma}\,\left(\mathbf{k}\cdot\gbf{\sigma}\right)\,\gbf{\psi}^{\pd}_{\mathbf{k}\sigma}\;, \label{eq:Hk4}
\end{align}
where ${\sigma =1,\ldots,N_f}$ denotes the fermion flavor, ${\gbf{\sigma}=(\gbf{\sigma}_x, \gbf{\sigma}_y)}$ is a vector of Pauli matrices and the spinor ${\gbf{\psi}^{\dagger}_{\mathbf{k}\sigma}=(a^{\dagger}_{\mathbf{k}\sigma},b^{\dagger}_{\mathbf{k}\sigma})}$ consists of the creation operators at individual orbitals with momentum $\mathbf{k}$. On a square lattice of linear system size ${L = 2N_k+1}$, where ${-N_k, -N_k+1, \ldots, N_k\in\mathbb{N}}$, with two orbitals per unit cell, the primitive vectors in the $x$- and $y$-directions and unit lattice constant, the Fourier transform ${a_{\mathbf{r}\sigma} = \sum_{\mathbf{k}} \E^{-\I\mathbf{k}\cdot\mathbf{r}} a_{\mathbf{k}\sigma}/L}$ then results in the bipartite real-space hopping Hamiltonian
\begin{eqnarray}
   H_{t\sigma} &=& -v_\text{F} \sum_{i=1}^{L^2} \left[\I \sum_{x=-N_k}^{N_k} t(x) \left(a_{i\sigma}^{\dagger} b_{i+x,\sigma}^{\phantom{\dagger}} - b_{i+x,\sigma}^{\dagger} a_{i\sigma} ^{\phantom{\dagger}}\right)\right. \nonumber\\
       &&+ \left. \sum_{y=-N_k}^{N_k} t(y) \left(a_{i\sigma}^{\dagger} b_{i+y,\sigma}^{\phantom{\dagger}} + b_{i+y,\sigma}^{\dagger} a_{i\sigma}^{\phantom{\dagger}}\right)\right] \;,
    \label{Ht}
\end{eqnarray}
where the amplitudes ${t(r)}$ correspond to the Fourier amplitudes from the transformation of \ceqn{eq:Hk4}, which decay ${\propto 1/r}$ in the thermodynamic limit (TDL) ${L\to\infty}$. Originally introduced for 3+1$d$ lattice field theories, SLAC fermions were constructed for odd ${L = 2N_k+1}$ finite-size systems \cite{Drell76}, where
\begin{align}
  t(r) = \begin{cases}
    0 & \text{if $r=0$,} \\
    \frac{(-1)^r \pi}{L}\frac{1}{\sin(r\pi/L)} & \text{otherwise.}
  \end{cases}
   \label{t_SLAC}
\end{align}
SLAC fermions on lattices with even linear dimension ${L = 2N_k}$, have the amplitudes \footnote{Contrary to \cref{Wang23b}, lattices with even linear dimension do not induce Gibbs ringing, but also exhibit the perfect linear dispersion once the appropriate Fourier transformation for has been performed as shown in \capp{ap:hoppingSLAC}.}
\begin{align}
  t(r) = \begin{cases}
    \frac{\pi}{L} & \text{if $r=0$,} \\
    \frac{(-1)^r \pi}{L}\left(\frac{\cos(r\pi/L)}{\sin(r\pi/L)} + \I\right) & \text{otherwise.}
  \end{cases}
   \label{t_SLAC_even}
\end{align}
By construction, this perfectly reproduces the relativistic dispersion with a $2N_f$-fold degeneracy at ${\mathbf{k}=(0,0)}$ for $N_f$ fermion flavors, such that the eigenvalues at all finite-size momenta obey ${\varepsilon_{\pm}(\mathbf{k})} = \pm v_\text{F} |\mathbf{k}|$, independently of the system size. 

\begin{figure}[t]
  \centering
  \includegraphics[width=0.8\columnwidth]{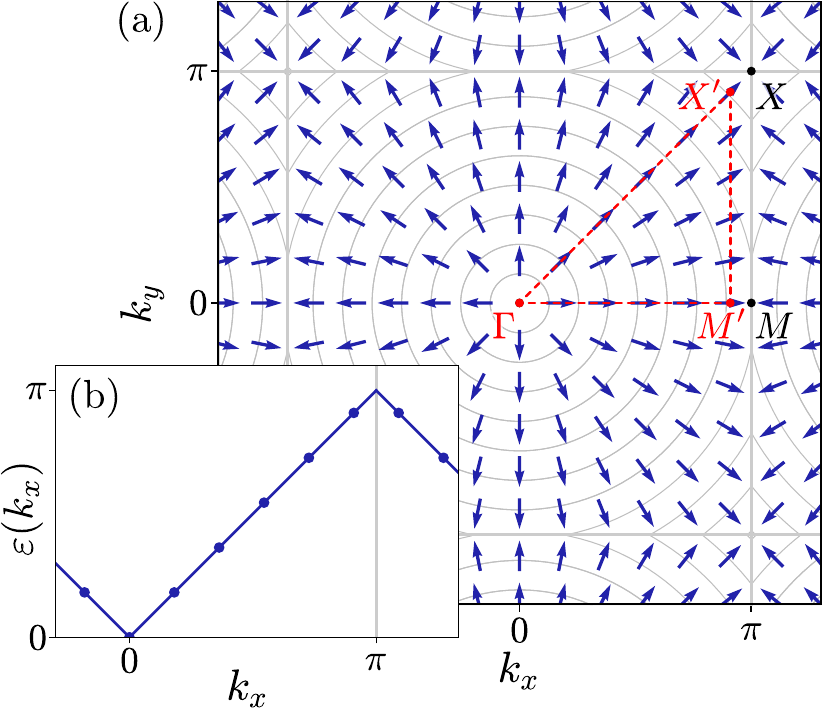}\\
  \caption{(a) Energy contours (light gray) and Berry flux pseudo spin map (blue arrows) for an $L=11$ lattice in momentum space. A closed contour around the Fermi point surface at the origin accumulates a flux of $\pi$ as can bee seen from the pseudo spin which sweeps twice the angle. The high symmetry path on a finite-size system is indicated by the red dashed line. The dispersion shown in (b) is discontinuous at the boundary of the Brillouin zone.}
  \label{fig:Berry}
\end{figure}
 
As required by time reversal symmetry, adiabatic motion of quasiparticles on the Fermi surface as their momentum goes around a loop $\mathcal{C}$ encompassing a band crossing point accumulates integer Berry flux, i.e., ${-\I\oint_{\mathcal{C}} \dd\mathbf{k}\cdot\langle\psi_{\mathbf{k}}|\nabla_{\mathbf{k}}|\psi_{\mathbf{k}}\rangle =n\pi}$, where $|\psi_{\mathbf{k}}\rangle$ is the Bloch wave function \cite{Haldane04}. The Dirac cone carries a Berry flux of $\pi$ as illustrated in the pseudo spin map ${\mathbf{k}\to \gbf{\tau}_{\mathbf{k}}/|\gbf{\tau}_{\mathbf{k}}|}$ of the eigenstates $|\psi_{\mathbf{k}}\rangle$ from momentum space to $S^2$ in \cfig{fig:Berry} (with degeneracy points removed) \cite{deGail12}. The correct Berry flux is essential for flux depending ordered states, such as quantum Hall order, but not necessarily for the existence of a phase transitions upon introduction of interactions. The dispersion, as well as the Berry flux are discontinuous at the Brillouin zone (BZ) boundary, which allows to place a single Dirac cone in the first BZ, but constitutes a potential source of singularities, or zeros in the interacting spectrum. Let us note, that any odd $L$ systems does not include the removable discontinuity at the boundary. As shown in \cfig{fig:Akw}, in the interacting system however, the dispersion across the periodic boundaries is renormalized and becomes smooth.

The number of poles of the Dirac operator $N$ is renormalization group relevant at the chiral transition \cite{Rosenstein93}. In order to make a proper comparison of models with SLAC fermions with local hopping lattices, such as the ${\pi}$-flux, or honeycomb lattice, the degeneracy of the fermions must be taken into account. While this does not affect the general properties of simulations with SLAC fermions, the critical properties will be different. The most basic Dirac cone in 2+1$d$ can be formulated for $N=2$ zero modes, i.e., spinless (spin-polarized) fermions with a single flavor $N_f$. While $N_f=1$ allows to study the chiral-Ising \cite{Tabatabaei22} and chiral-XY \cite{Li18} quantum phase transitions at least two flavors per orbital are required to realize a system where SU(2) symmetry can be spontaneously broken upon the introduction of, e.g, a local Coulomb repulsion. The ${2\times 2}$ representation of \ceqn{Ht} can be easily generalized by increasing the number of flavors. For ${N=4, N_f=2}$ this has been realized in \cite{Lang19,Xu21} to investigate the chiral-Heisenberg transition with a single Dirac cone. For ${N_f > 2}$, in the adjoint representation, which is suitable for sign-problem-free AF-QMC simulations \cite{Lang13, Assaad05, Li16, Li19}, this however increases the symmetry of the fermions to ${\mathrm{SU}(N_f)\times\mathrm{SU}(N_f)}$ and a naive generalization of the Hubbard interaction would tend to break the $\mathrm{SU}(N_f)$ symmetry, rather than the targeted SU(2) flavor symmetry. In order to study the chiral-Heisenberg transition, e.g., with ${N=8}$ either the interaction terms must be chosen, such that solely the SU(2) subgroup is spontaneously broken. Alternatively, one may set the number of orbitals ${N_{\ell}=4}$ per site while keeping $N_f=2$: The ${4\times 4}$ Hamiltonian
\begin{align}
   H_{\mathbf{k}\sigma} = - v_\text{F} \sum_{\mathbf{k}}\gbf{\psi}^{\dagger}_{\mathbf{k}\sigma}\,\left(\mathbf{k}\cdot\gbf{\Gamma}\right)\,\gbf{\psi}^{\pd}_{\mathbf{k}\sigma}\;,
	\label{eq:Hk8}
\end{align}
with ${\gbf{\Gamma} = (\gbf{\sigma}_0\otimes\gbf{\sigma}_x, \gbf{\sigma}_0\otimes\gbf{\sigma}_y)}$ and the 4-spinor ${\gbf{\psi}^{\dagger}_{\sigma}}$ 
then is the ${N=8}$ generalization of \ceqn{eq:Hk4}. This formulation results two congruent Dirac cones with $N$ zero modes in total, Berry flux $\pi$ each and can be augmented by the Hubbard interaction
\begin{align}
	H = \sum_{\sigma=1}^{N_f} H_{\mathbf{k}\sigma} + \frac{U}{N_f} \sum_{i=1}^{L^2}\sum_{\ell=1}^{N_{\ell}} \left[\sum_{\sigma=1}^{N_f} c_{i\ell\sigma}^{\dagger} c_{\ell i\sigma}^{\phantom{\dagger}} - 1\right]^2 \;,
	\label{eq:HtU}
\end{align}
where $c_{i\ell}^{\dagger}$ creates a fermion at site $i$ in orbital $\ell $ and which is expected to break $\mathrm{SU}(2)$ spin (flavor) symmetry at strong coupling $U/t\gg 1$. In the following we choose ${v_\text{F} = 1}$ as a unit of energy.

\subsection{Method}

We employ a projector axillary-field determinantal QMC scheme by which we obtain zero temperature momentum and spin resolved single-particle Green's function in imaginary time for finite lattices with ${N L^2/2}$ sites for SLAC fermions and $2L^2$ sites for the honeycomb lattice with periodic boundary conditions. We stochastically project out the ground state from the half-filled slater determinant of the non-interacting system using a projection length ${2\Theta=70}$ and imaginary-time step ${\Delta\tau = 0.1}$. An imaginary-time evolution length ${\tau = 30}$ has been used to obtain the Green's function. The Hamiltonian of \ceqn{eq:HtU} is free from a negative sign problem in QMC simulations \cite{Li16, Li19} and the simulation with SLAC fermions allows for local fast updates. Details of the algorithm can be found in Ref.~\cite{Assaad08}.

A few technical drawbacks of the SLAC formulation compared to simulations on the honeycomb should be noted: (i) The numerical costs in AF-QMC simulations for the $N=8$ SLAC formulation with its four orbitals per unit cell scale as $O(64 L^6)$, which is higher than the $O(8 L^6)$ scaling of the honeycomb. As we will show though, the significantly smaller finite-size effects make the SLAC formulation arguably still superior to the honeycomb implementation. This overhead could be overcome by using the antisymmetic SU($N$) representation on the half-filled lattice, with an interaction term which solely breaks the targeted SU(2) symmetry, which we leave to future investigations. (ii) The non-local nature of the hopping term decreases the benefit employing a checkerboard breakup of the hopping matrix exponential. (iii) While we will show that the implementation with four fermion interactions works well, the combination of SLAC fermions with dynamical gauge fields constitutes not only a technical challenge, but also raises the questions with respect to the locality and renormalizability of such iteracting systems as described below. (iv) For even $L$ the last term in \ceqn{t_SLAC_even} corresponds to a constant amplitude, long range oscillating hopping, which will dynamically induce a homogeneous antiferromagnetic (AFM) magnetic field proportional to ${\pi/L^2}$ as discussed below. While it vanishes with system size, this symmetry breaking field will results in different finite-size behavior compared to odd $L$ systems.

\section{Locality \& dynamically induced interactions}\label{locality}

In this manuscript we simulate the Hamiltonian lattice model in \ceqn{eq:HtU}, rather than the GNY lattice field theory itself. We have to be mindful of the apparent long range structure of the Dirac operator in real space, which stands in stark contrast to the inherent locality of physical quantum field theories, but allows to avoid the Nielsen-Ninomiya theorem \cite{Nielsen81a,Nielsen81b,Nielsen81c} on finite-size systems and to place a single Dirac cone onto the lattice \footnote{Note that in odd spacetime dimensions considered in this work chiral symmetry is replaced by parity symmetry \cite{Winkler15}.}. The real-space hopping however is not equivalent to a genuine long range coupling, as the hopping only runs along the major axes. This couples a given site only to a finite fraction of the total volume. While this does not exclude possible spurious long-range effects, it suggests their potential influence scales intensively and decreases with increasing system size and dimensionality.   

SLAC fermions have been successfully employed in simulations without gauge fields, such as supersymmetric Yukawa, Wess-Zumino and Thirring models \cite{Karsten79,Kirchberg05,Bergner08,Kaestner08,Bergner10,Flore12,Wellegehausen17}. The exactly known critical exponents of the supersymmetric ${N=2}$ chiral-XY transition were accurately reproduced by spinless SLAC fermions \cite{Zerf17,Li18}. In addition we point to the good agreement between the chiral Ising critical properties from simulations with a single Dirac cone \cite{Tabatabaei22} and the recent results from conformal bootstrap \cite{Erramilli23}. While the chiral Heisenberg transition could be exceptional, the estimates from the other chiral transitions using SLAC fermions are encouraging and do not exhibit pathological signatures.

In this section we discuss the effects of the dynamically induced AFM long-range spin interaction as a result of long-range hopping and present evidence that in our simulations we obtain well-defined phases in the absence of spurious zero modes or other artifacts, which would render the system non-renormalizable. 

\subsection{Mermin-Wagner criterion \& finite temperature}

The Hamiltonian lattice model in \ceqn{eq:HtU} and SLAC fermions are by construction effective low energy models akin to an effective field theory. As such their application is confined by energy scales and dimensionality \cite{Drell76}. Indiscriminate application of the SLAC Hamiltonian in a regime, dimensionality, or limit it is not intended for, can lead to physical scenarios which are not compatible with the original local Hamiltonian the effective Hamiltonian has been modelled on. This has been demonstrated in \cite{Gebhard92,Gebhard94,Wang23b} where the effects of long range hopping result in the spontaneous breaking of continuous symmetries in one dimension. In 2$d$ at finite temperature the authors of \cite{DaLiao22} report an interaction induced finite temperature phase transition in a SLAC-Hubbard model, which is absent in the purely short-range $\pi$-flux Hubbard model on the square lattice. While at odds with the physics of the short-range model, which forbids the spontaneous breaking of a continuous symmetry, the model with SLAC fermions does not violate the Hohenberg-Mermin-Wagner (HMW) theorem \cite{Mermin66, Hohenberg67}, since it does not apply. 

Despite claims of a break down, or violation of the HMW theorem in models with fully connected long range interactions \cite{DaLiao22, Zhao23}, for spin-couplings ${J \sim r^{-\alpha}}$ with ${\alpha < 4}$ the conditions of the theorem are not satisfied: The theorem considers a magnetically ordered 2$d$ spin system at finite temperature and the energy necessary to create a magnetically disordered droplet of volume $V\sim L^2$ with linear size $L$, which is of the order $\sum_{r\in V} r^2 J(r)$, \cite{Kunz76}. If this energy is bounded, independently of the size $V$, the disorder droplets are probable and destroy the magnetic order. For fully (all-to-all) connected long range interactions of the kind ${J \sim r^{-\alpha}}$, where ${\mathbf{r} = (x,y)}$, the droplet energy is at worst 
\begin{align}
	\sum_{r\in V} r^2 J(r) = \!\!\!\sum_{x,y=-L/2}^{L/2} r^2 r^{-\alpha}\; \sim\; L^{4-\alpha}\;.
\end{align}
For ${\alpha < 4}$, the energy of the droplet thus grows as a power law, rendering them unlikely and order prevails. The consequence is that the emergence of order which breaks a continuous symmetry is perfectly cromulent in ${2d}$ at finite temperature as long as the spin exchange coupling decays slower than ${J\sim r^{-4}}$. Shorter ranged couplings, where ${\alpha > 4}$, result in a bounded droplet energy leading to disorder at finite temperature. 

The SLAC formalism, can be imagined as an 2$d$ bilayer system with all-to-${2L}$ connectivity along the major axis, such that the connectivity scales sub-volume. 
The SLAC hopping matrix elements ${t \sim 1/r}$, generate to leading order the effective AFM exchange coupling  ${J \sim t^2/U \sim 1/U r^2}$. In analogy with the example above, at strong coupling, for major-axis connectivity the disorder droplet energy becomes
\begin{align}
	\sum_{r\in V} r^2 J(r) = \!\!\!\sum_{x=-L/2}^{L/2} r^2 r^{-\alpha} + \!\!\!\sum_{y=-L/2}^{L/2} r^2 r^{-\alpha}\; \sim\; L^{3-\alpha}\;,
\end{align}
where ${\alpha < 3}$ suffices to allow for the energy to grow, or to allow for the spontaneous breaking of continuous symmetry in ${2d}$ at finite temperature. Consequently, for ${J \sim 1/r^2}$ the HMW theorem does not apply, at least within this leading order approximation. Higher order processes will also include spin exchange contributions off the major-axis, but these will likely decay significantly faster than along the major axis. The consequence is, that long-range order can occur within a SLAC-Hubbard model due to the spontaneous breaking of a continuous symmetry in $d<3$ even at finite temperature without violating the HMW theorem. The finite temperature transition is not to be expected in the chiral Heisenberg universality class due to the small exponent ${\alpha=2}$ and the limited dimensionality.

\subsection{Zero temperature}

At temperature ${T=0}$ the HMW theorem certainly no longer restricts continuous symmetry breaking, as we enter a dimensionality ${d+z}$, where $z$ denotes the dynamic scaling exponent, relating the scaling between the energy and momentum. In particular, Lorentz- (or more specifically Poincar{\'e}-) invariance emerges with ${z=1}$ at the critical point in local and in SLAC models \cite{Roy16,Lang19,Tabatabaei22}, which can be identified from the scaling of the single particle gap ${\Delta_\text{sp}\sim L^{-z}}$ at the critical coupling in the inset of \cfig{fig:m2_gaps}. However, the dynamically induced long-range super-exchange, if not forbidden by symmetries, will still be present. Here, we investigate its influence and the properties of the Dirac semi-metal (DSM) at weak coupling, as well as the strong coupling AFM and the critical properties.

\subsubsection{Weak coupling Dirac semi-metal}

\begin{figure}[t]
  \centering
  \includegraphics[width=\columnwidth]{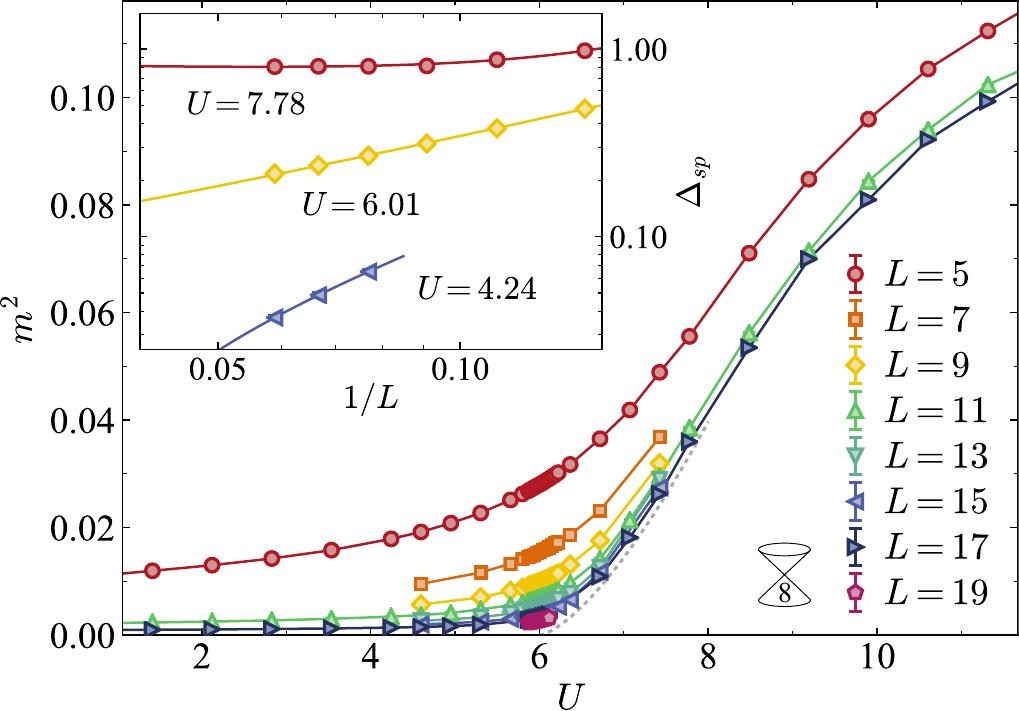}\\
  \caption{The model shows two well defined phases: a gapless DSM regime where the order parameter ($m^2$) vanishes and a gapped, AFM ordered phase where the order parameter acquires finite values. Inset: the finite-size extrapolations of the single particle gap $\Delta_\text{sp}$ below, at and above the critical coupling ${U_c\approx 6}$.
  \label{fig:m2_gaps}}
\end{figure}

We observe a stable DSM phase up to the critical point, beyond which AFM order emerges and the fermions acquire mass. The emergence of long-range AFM order can be tracked by measuring the spin structure factor ${S_\text{AFM}(\mathbf{k}) \equiv \sum_{\mathbf{r}} \E^{\I\mathbf{k}\cdot\mathbf{r}} \langle\mathbf{S}(\mathbf{r})\cdot\mathbf{S}(\mathbf{0})\rangle/L^2}$, where the spin ${\mathbf{S}_{\mathbf{r}} = \frac{1}{2}a_{\mathbf{r}\alpha}^{\dagger}\gbf{\sigma}_{\alpha\beta} a_{\mathbf{r}\beta}}$ at position $\mathbf{r}$, orbital $a$, and $\gbf{\sigma}$ denotes the vector of the three Pauli matrices. The squared magnetization then is obtained from ${m^2 = S_\text{AFM}(\mathbf{k}=0)/L^2}$. Figure~\ref{fig:m2_gaps} shows the evolution of the finite-size order parameter as a function of the Coulomb repulsion, which indicates the quantum phase transition to be located at ${U_c\approx 6}$. The inset shows the finite-size extrapolation of the single-particle gap $\Delta_\text{sp}(\mathbf{k})$ obtained from a fit to the asymptotic long imaginary-time behavior of the single particle Green's function ${G(\mathbf{k},\tau) = \langle a_{\mathbf{k}}^{\dagger}(\tau)\, a_{\mathbf{k}}^{\phantom{\dagger}}(0)\rangle \propto \exp[-\tau\Delta_\text{sp}(\mathbf{k})]}$. The gap vanishes faster than ${1/L}$ consistent with simulations of local models on a torus \cite{Lang19,Seki19,Schuler21} below the critical coupling ${U\le U_c}$, such that the Dirac fermions remain massless in the TDL and the gap opens with the onset of order. We detect no anisotropy in the single particle and spin Green's functions, all supporting a stable semi-metal fixed point. Potential long-range spin fluctuations as a consequence of dynamically induced super-exchange mediated by the long range-hopping apparently do not destabilize the DSM. 

\begin{figure}[t]
  \centering
  \includegraphics[width=\columnwidth]{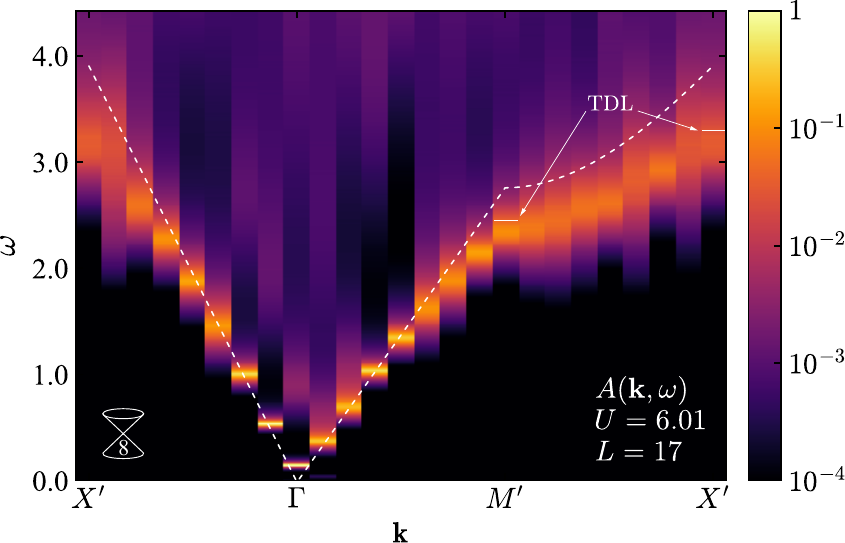}\\
  \caption{The single-particle spectral function ${A(\mathbf{k},\omega)}$ close to criticality for $L=17$ along the path in the Brillouin zone indicated in \cfig{fig:Berry}. The non-interacting dispersion is indicated by the dashed line. The lowest excitation energy in the TDL are indicated for ${\mathbf{k}=M'}$, $X'$.}
  \label{fig:Akw}
\end{figure}

The locality of the Dirac operator is recovered in the infinite volume limit for most of the BZ, but the singular boundary \cite{Bodwin88,Campos02,Kirchberg05}. We thus monitor the dominant single particle channel and AFM correlations for interfering low-energy modes induced by interactions, in particular close to the boundary of the BZ. Such excitations (doublers) could potentially alter the degeneracy of the fermions at the Fermi level and consequently the universality class of the quantum phase transition. The single-particle spectral function ${A(\mathbf{k},\omega) = -{\rm Im}\, G(\mathbf{k},\omega)/\pi}$ close to the critical point as obtained from stochastic analytical continuation \cite{Sandvik98,Beach04} is presented in \cfig{fig:Akw} for momenta along the path in the BZ sketched in \cfig{fig:Berry}(a). The dispersion of the non-interacting system is indicated by the white dashed line. The excitation gap at $\Gamma$ is the one most affected by finite-size corrections, yet scales predominately as ${\Delta_\text{sp}\propto L^{-z}}$. The finite-size corrections of the single-particle excitations away from vanishing momentum are significantly smaller leading to a scaling faster than $L^{-1}$  \cite{Hesselmann19,Schuler21}. Note that the quasiparticle weight $Z$ (the intensity in \cfig{fig:Akw}) actually vanishes at ${\mathbf{k}\to\Gamma}$ at the critical point as ${Z\sim (U-U_c)^{-z\nu}}$ in the TDL \cite{Lang19,Herbut09b}. The discontinuous dispersion at the BZ boundary appears to be smoothed upon the introduction of interactions and the single particle excitations close to the boundary of the BZ converge to a finite values at rather high energies. We indicate the finite-size extrapolated values of the excitations at the momenta ${M'=(\pi-2\pi/L,0)}$ and ${X'=(\pi-2\pi/L,\pi-2\pi/L)}$ closest to the Brillouin zone boundary. No additional zero modes introduced by correlations as reflected in the interacting single-particle spectrum, nor the spin spectrum as discussed below at criticality.

\begin{figure}[tp]
  \centering
  \includegraphics[width=\columnwidth]{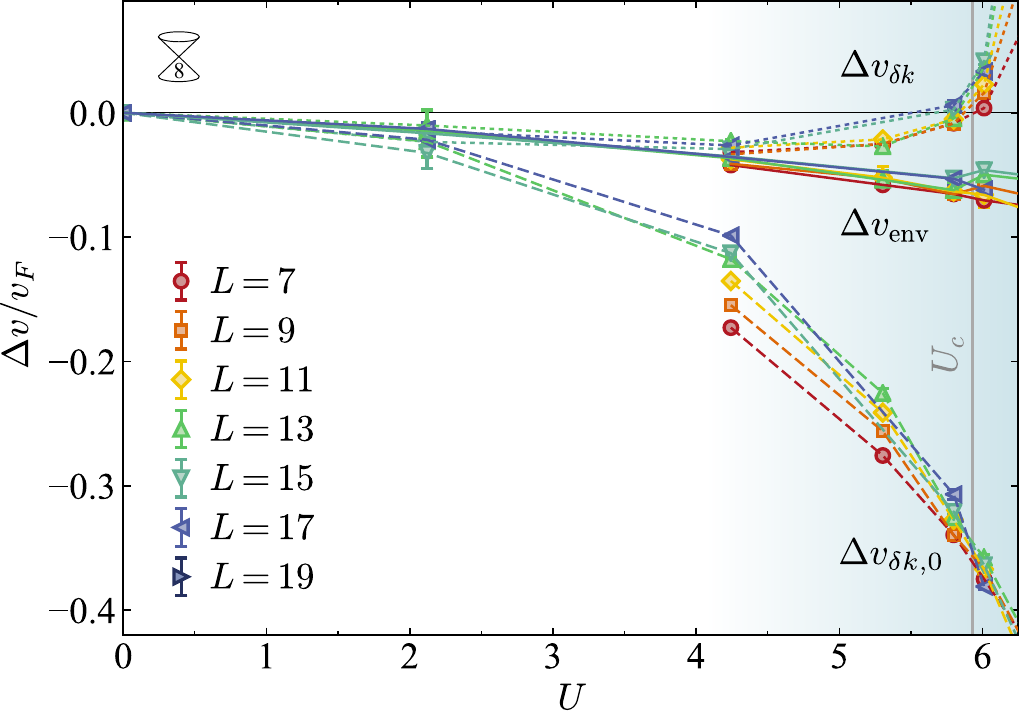}\\
  \caption{Estimators for the renormalization of the Fermi velocity based on the finite-size energies at the smallest lattice momentum $\Delta(\delta\mathbf{k})$, the relativistic envelope of ${\Delta(\mathbf{k})}$ and the Dirac point $\Delta(\mathbf{0})$, . The finite-size  show minor renormalization of the Fermi velocity.
  \label{fig:vF}}
\end{figure}

The envelope of the dispersion close to criticality suggests there is only weak renormalization of the Fermi velocity. This is in accordance with results from appropriately chosen estimators on the honeycomb and other lattices, as well as for ${N=4}$ \cite{Hesselmann19,Liu19,Lang19}. In the absence of the knowledge of the level structure from conformal field theory we cannot determine the \textit{true} Fermi velocity at the critical point \cite{Schuler16,Whitsitt17,Schuler21}. We can however track the Fermi velocity approaching the critical point from the noninteracting DSM fixed point, where the \textit{speed of light} $v_F$ is known, as shown in \cfig{fig:vF}. 
Here we extract the fermionic excitations velocity on the torus from fits of the relativistic energy momentum relation 
to the envelope of the excitations gaps $\Delta(\mathbf{k})$ in the range ${k < 1.3}$, such that for different system sizes
\begin{equation}
	v_\text{env} = \sqrt{\frac{\Delta^2(\mathbf{k})-\Delta^2(\mathbf{0})}{k^2}}\;,
	\label{eq:venv}
\end{equation}
as well as from the complementary estimator including the excitation gap closest to the Dirac point ${\delta\mathbf{k} = (2\pi/L,0)}$ only
\begin{equation}
	v_{\delta\mathbf{k}} = \frac{\Delta_{sp}(\delta\mathbf{k})}{\delta k}\;.
	\label{eq:vdk}
\end{equation}
While the definition for $v_\text{env}$ includes the strongly finite-size affected Dirac point, the use of SLAC fermions allows to reliably extract the excitations velocity from momenta where the dispersion is approximately linear. This is in contrast to the definition
\begin{equation}
	v_{\delta\mathbf{k},0} = \frac{\Delta(\delta\mathbf{k})-\Delta(\mathbf{0})}{\delta k}\;,
	\label{eq:vdk0}
\end{equation}
which includes the Dirac point and can lead to inconsistent estimates of the velocity renormalization \cite{Hesselmann19}. 

Despite the non-local nature of the SLAC derivative we observe a well defined DSM phase in the absence of pathological signatures in our simulations and small renormalization of the Fermi velocity.

\subsubsection{Strong coupling AFM}

\begin{figure}[t]
  \centering
  \includegraphics[width=\columnwidth]{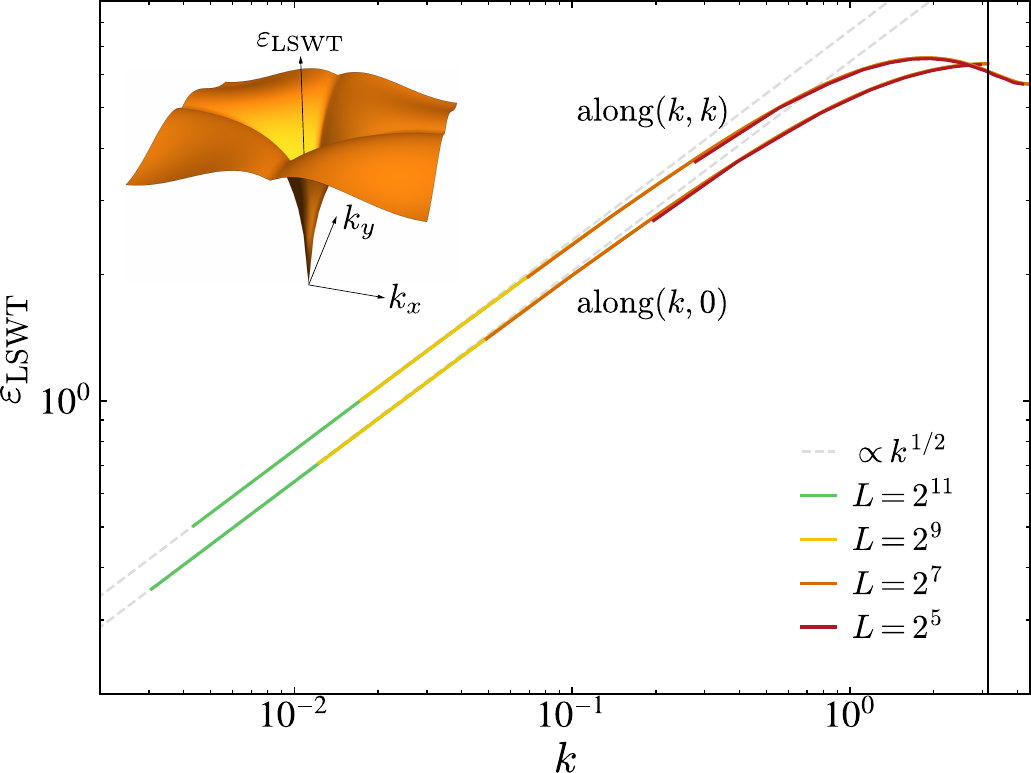}\\
  \caption{The spin-wave dispersion from LSWT (inset) for AFM couplings ${\sim 1/r^2}$ along the major axis shows a distinct anisotropy. The dispersion opens as ${\sqrt{k}}$ for small momenta in all directions, here shown along the paths $(0,0)$ to ${(\pi,0)}$ and ${(\pi,\pi)}$.
  \label{fig:LSQT_powerlaw}}
\end{figure}

\begin{figure}[t]
  \centering
  \includegraphics[width=0.95\columnwidth]{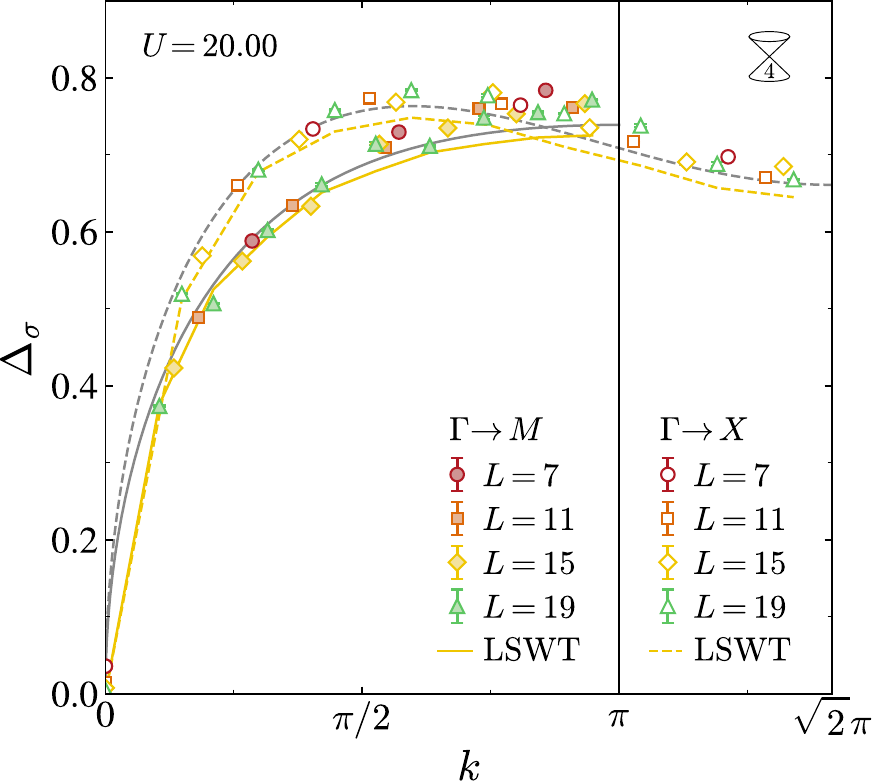}\\
  \caption{The momentum dependence of the QMC spin gap $\Delta_{\sigma}(k)$ at strong coupling ${U=20}$ along the paths $(0,0)$ to ${(\pi,0)}$ and ${(\pi,\pi)}$ is well reproduced by the finite-size LSWT (orange ${L=19}$, gray ${L\to\infty}$).
  \label{fig:LSQT_disp}}
\end{figure}

At strong coupling ${U\gg v_F}$ the system spontaneously orders antiferromagnetically, driven by an effective super-exchange coupling, which results from predominantly second order hopping processes as ${J\sim t^2/U \sim 1/U r^2}$. As the fermions are gapped out we can approximate the mostly localized ground state as Heisenberg ${S=1/2}$ antiferromagnetic via the Holstein-Primakoff transformation to linear order \cite{Holstein40}. The resulting bosonic Hamiltonian in the basis ${(a_\mathbf{k}^{\dagger}, b_\mathbf{k}^{\pd})}$, where the two orbitals are associated with the sublattices at momentum $\mathbf{k}$, has the form
\begin{align}
	H(\mathbf{k})
	&= \left(a_\mathbf{k}^{\dagger}, b_\mathbf{k}^{\pd}\right)
      \begin{pmatrix}
      	\lambda_\mathbf{k} & J_\mathbf{k} \\ 
      	J_\mathbf{k} & \lambda_\mathbf{k}\\ 
      \end{pmatrix}
	  \begin{pmatrix}a_\mathbf{k}^{\pd}\\ b_\mathbf{k}^{\dagger}\end{pmatrix} - \lambda_\mathbf{k}\;,
	  \label{LSWT_H}
\end{align}
up to constants. Here $J_\mathbf{k}$ is the Fourier transformed coupling between the sublattices
\begin{align}
	J_{\mathbf{k}}
	&= \sum_{x=-L/2}^{L/2}\!\!\!\!\!{}^{'}J(|x|)\,\E^{-\I k_x x} + \sum_{y=-L/2}^{L/2}\!\!\!\!\!{}^{'}J(|y|)\,\E^{-\I k_y y} \nonumber\\
	&= \frac{2\pi^2}{L^2}\left[\sum_{r=1}^{L/2}\frac{\cos(k_x r)}{\sin(r\pi/L)^2} + \sum_{r=1}^{L/2}\frac{\cos(k_y r)}{\sin(r\pi/L)^2}\right] \;,
	  \label{LSWT_J}
\end{align}
where we have used the explicit finite-size expression for ${t(r)}$ from \ceqn{t_SLAC}. The Hamiltonian (\ref{LSWT_H}) may be diagonalized by a Bogoliubov transformation, which yields the eigenvalues
\begin{align}
	\varepsilon_\mathbf{k} = \sqrt{\lambda_\mathbf{k}^2-J_\mathbf{k}^2}\;.
\end{align}
To guarantee $H$ to be positive definite,
\begin{align}
	\lambda_\mathbf{k} = J_{\mathbf{k}=\mathbf{0}} = \frac{4\pi^2}{L^2}\sum_{r=1}^{L/2}\frac{1}{\sin(r\pi/L)^2}\nonumber\\ \stackrel{L\to\infty}{=} 4\sum_{r=1}^{\infty}\frac{1}{r^2}  = 4\zeta(2) = \frac{2\pi}{3}\;.
\end{align}
The spin-wave dispersion in the thermodynamic limit is then given by
\begin{align}
	\varepsilon_\text{LSWT}(\mathbf{k}) = \sqrt{\left(\frac{2\pi}{3}\right)^2-J_\mathbf{k}^2}\;,
    \label{LSWT_ek}
\end{align}
whereas for finite-size systems we use ${\varepsilon_\text{LSWT}(\mathbf{k}) = \sqrt{J_{\mathbf{0}}^2-J_\mathbf{k}^2}}$. While retaining all lattice symmetries, the resulting dispersion is anisotropic as a consequence of the major axis coupling and shown in the inset of \cfig{fig:LSQT_powerlaw}. In contrast, in fully connected long range models the magnon dispersion is isotropic. For small momenta the dispersion follows $\sqrt{k}$ similar to fully connected long range models \cite{Diessel22}, but for the major axis anisotropy. At strong coupling \ceqn{LSWT_ek} approximates the true spin wave dispersion. This is shown for ${U=20}$,  ${N=4}$ along the cuts ${(k,0)}$ and ${(k,k)}$ in \cfig{fig:LSQT_disp}, where we extracted the spin gap $\Delta_{\sigma}$ obtained from a fit to the asymptotic long imaginary-time behavior of the unequal-time spin-spin correlation function ${S(\mathbf{k},\tau) = \langle S_{\mathbf{k}}^{\dagger}(\tau)\, S_{\mathbf{k}}^{\phantom{\dagger}}(0)\rangle \propto \exp[-\tau\Delta_{\sigma}(\mathbf{k})]}$.

In the SU(2) spin symmetry broken phase one expects the emergence of Goldstone modes, which can be tested by the extrapolation ${\Delta_{\sigma}(\mathbf{k}\to 0^+)\to 0}$. Due to the nonlinear $\sqrt{k}$-form of the dispersion at small momenta, a naive extrapolation will significantly overestimate the gap ${\Delta_{\sigma}(0^+)}$ in finite-size systems with limited momentum resolution. In order to reduce the finite-size effects in the extrapolation, we rescale the finite-energy spectra at momentum $\mathbf{k}$ as
\begin{align}
	\Delta_{\sigma}(\mathbf{k})\to \hat{\Delta}_{\sigma}(\mathbf{k}) = \frac{v_{\phi} |\mathbf{k}|}{\varepsilon_\text{LSWT}(\mathbf{k})}\, \Delta_{\sigma}(\mathbf{k})\;,
\end{align}
where $v_{\phi}$ denotes the spin-wave velocity fitted to the data. This \textit{unwarping} enables us to remove the leading non-linear behavior from the finite-size data, yet retains the correct limit for vanishing momentum \cite{Schuler21}. The then approximately linear dispersion allows for the extrapolation of the spin gaps to zero, which is consistent with the emergence of gapless Goldstone modes. While anomalous in their $\sqrt{k}$-dispersion, compared to a fully connected effective long-range Heisenberg model, the behavior also implies that the dynamically generated super-exchange coupling decays significantly faster than would be required for a gapped Goldstone mode \cite{Diessel22} which has been claimed for the ${N=4}$ SLAC-Hubbard model in \cite{DaLiao22}.


\begin{figure}[t]
  \centering
  \includegraphics[width=0.92\columnwidth]{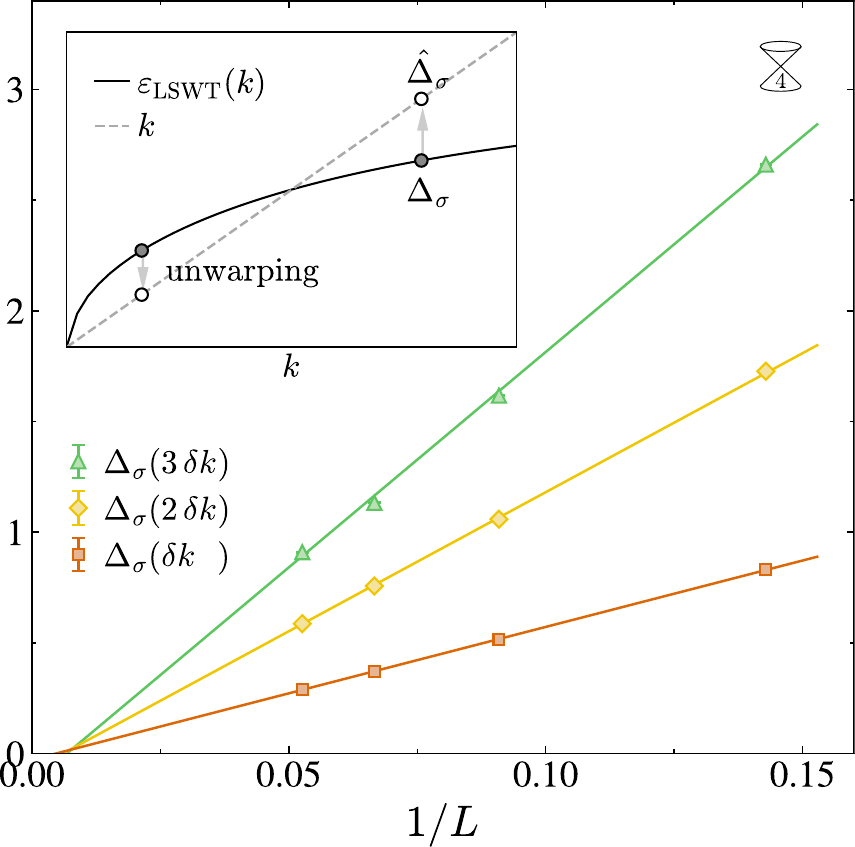}\\
  \caption{finite-size extrapolation the unwarped QMC data at strong coupling ${U=20}$ (c.f. inset) via the LSWT spin-wave dispersion ${\varepsilon_\text{LSWT}}$ to remove the leading nonlinear behavior. The linear finite-size extrapolation of the smallest unwarped spin gaps ${\hat{\Delta}_{\sigma}(k\to 0)}$ to zero provides evidence for gapless, anomalous Goldstone modes.
  \label{fig:LSQT_unwarped}}
\end{figure}

The dynamically induced super-exchange along the major axis results in an anisotropic spin-wave dispersion accompanied by anomalous, yet gapless Goldstone modes in the AFM insulator. Despite the anisotropy, the AFM order still breaks the same SU(2) spin rotational symmetry while retaining the same lattice symmetries as in local interacting fermion models. We therefore argue that we are considering the same strong coupling fixed point but for specific details of the spin correlations. Noticeably, we find no magnon excitations below the light cone at small momenta. This implies that the dynamical super-exchange does not induce instantaneous non-causal interactions, or only with vanishingly small amplitudes.

\subsubsection{Criticality}

\begin{figure}[t]
  \centering
  \includegraphics[width=0.98\columnwidth]{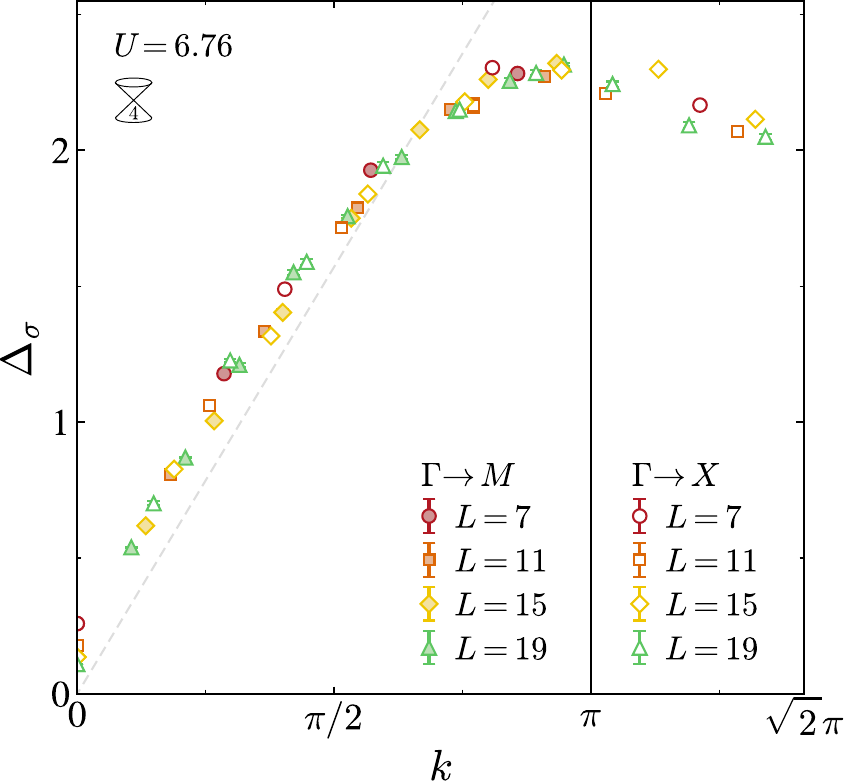}\\
  \caption{Close to the critical point the spin excitations appear isotropic and approximately reproduce the (non-interacting) Fermi velocity (dashed gray line, also cf. \cfig{fig:vF}), albeit with an unclear finite-size dependence for finite momenta.}
  \label{gap_SPIN_critU_isotropy}
\end{figure}

The question remains whether the effective long-range coupling alters the properties at the critical point which then would no longer be representative of the ${N=8}$ chiral Heisenberg universality class. First, we observe finite-size single particle and spin gaps that vanish in the TDL (c.f. inset in \cfig{fig:m2_gaps}, \cfig{fig:LSQT_unwarped} and \cfig{gap_SPIN_critU_isotropy}) consistent with gapless excitations in the two relevant sectors. Across criticality we further expect the fermionic Dirac spectrum to be replaced by its bosonic counterpart, such that the primary excitations share the same velocity \cite{Herbut09a,Roy11,Roy16}. In \cfig{gap_SPIN_critU_isotropy} we can see that at criticality and in contrast to \cfig{fig:LSQT_disp} at strong coupling, the spin wave dispersion is isotropic, with an excitation velocity approaching the non-interacting Fermi velocity as indicated by the gray dashed line. This also provides an independent indicator that there is little to no renormalization of the Fermi velocity. Where the finite-size dependence of the spin gaps at zero momentum is clear, for finite momenta this is less the case. Whether the shift to higher energies off the dashed line indicating the non-interacting Fermi velocity is a finite-size effect or a persistent remnant of the dynamically induced long-range spin-exchange remains open. Again, we find no evidence for super-luminal magnon excitations. The recovered isotropy and similar velocities of the spin excitations compared to the fermionic single particle spectrum are in accordance with the expectations for the GNY-type transition.

An additional worry would be, whether the dynamical long-range interactions preclude the fundamental necessary property of renormalizability. Evidence for the renormalizability is provided by dimensionless quantities, which remain finite with increasing system size, while the correlation length diverges as one approaches the critical point. The dimensionless ratios $R_{m^2}$ we introduce in the following \csec{FSS} fulfill this very property. Furthermore, we successfully employ a finite-size scaling ansatz and are able to fit and collapse the finite-size data accordingly. The working finite-size scaling points towards the fact that the model is (at least) non-perturbatively renormalizable. We observe no pathological signatures up to the simulated lattice sizes. Note that in our investigation we do not need to take the continuum limit, but only the infinite volume limit, such that the lattice provides a natural, persistent ultraviolet cutoff.

The dynamically induced long-range interactions does not appear to alter the critical properties. Indeed, long-range simulations of fermions on the honeycomb lattice subject to Coulomb interactions with a long-range part ${V(r)\sim 1/r}$ also have provided data that are approximately consistent with chiral Heisenberg Gross-Neveu universality \cite{Hohenadler14,Tang18,Hesselmann19} and are supported by long-range GNY models studied in \cite{Chai22}.

\section{Critical scaling properties\label{FSS}}

\subsection{Proximity to asymptotic scaling\label{Rm2}}

It is arguable that many numerical investigations of critical phenomena, even when of the highest precision, fall short of the asymptotic scaling regime. This is the case for simulations that are not performed close enough to the critical point, or on insufficiently large lattices. \textit{Close}, or \textit{large enough} is determined by the crossover scale where the finite-size spectrum coincides with the chiral Heisenberg conformal field theory provided the linear extent of the simulation torus satisfies 
$L \gg |U-U_c|^{-\nu}$ \cite{Schuler21}. In order to determine if a certain lattice model is closer than the another, one can directly compare complementary quantities, in order to identify favorable finite-size scaling behavior. 

Many different dimensionless quantities can be constructed, such that the optimal choice of operators, i.e., the one which picks up the lowest order scaling corrections is ambiguous. Here we consider as RG invariant quantity the correlation ratio (CR)
\begin{align}
   R_{C}^{(n_1,n_2)} = 1 - \frac{C(\mathbf{Q} + n_1 \mathbf{b}_1 + n_2 \mathbf{b}_2)}{C(\mathbf{Q})}\;,
\end{align}
where $\mathbf{b}_1$, $\mathbf{b}_2$ denote the reciprocal lattice vectors and $\mathbf{Q}$ is the momentum at which order develops across the transition \cite{Kaul15,Pujari16}. The correlation function $C$ must be associated with the order parameter of the phase transition. Here, the correlation function corresponds to the spin structure factor ${S_\text{AFM}(\mathbf{Q})/L^2 = m^2}$ and we will denote the CR as $R_{m^2}$. The critical point can be inferred from the intersections of different pairs of system sizes where ${R_{m^2}(u,L) = R_{m^2}(u,aL+b)}$. Here ${a,b\in\mathbb{N}}$ with a typical choice of ${a=2}$ in order to express the dominant scaling behavior and ${b=0,1}$ to account for commensurate lattice sizes. In the TDL at the critical coupling the CR approaches ${\lim_{L\to\infty} R_{m^2}(0,L) = R_{m^2}^*}$ \cite{Shao16,Nvsen18}. 

\begin{figure}[t]
  \centering
  \includegraphics[width=\columnwidth]{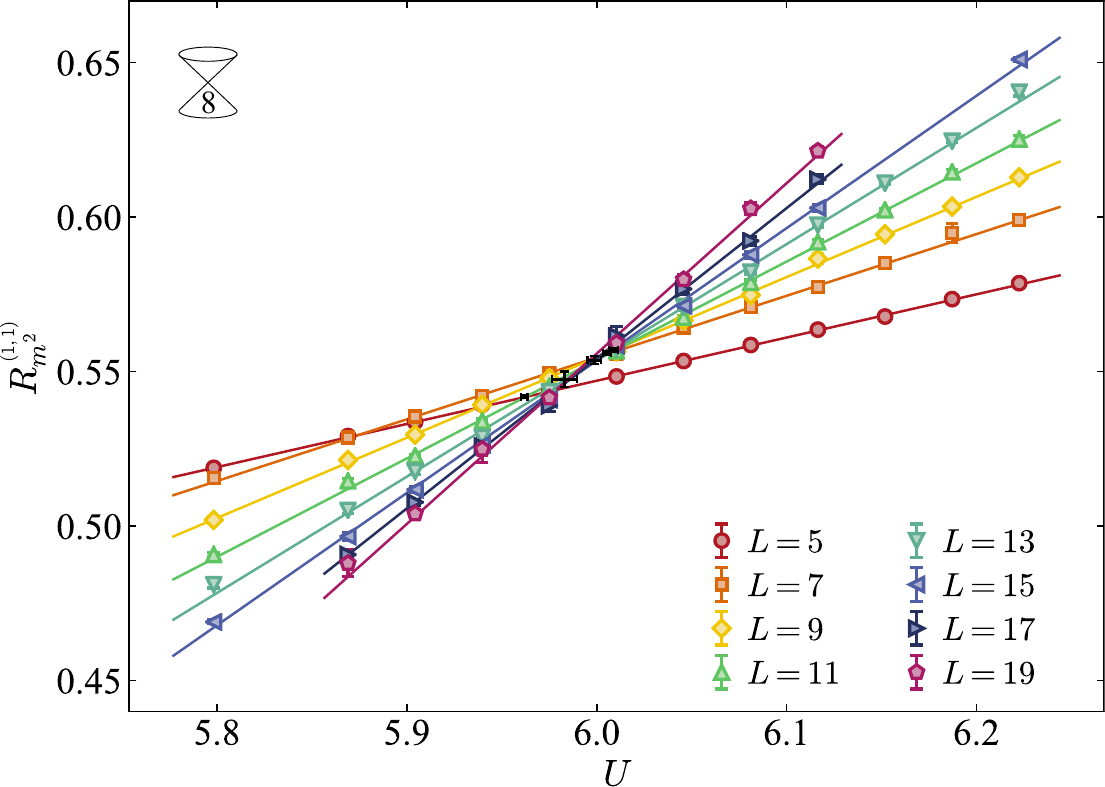}\vspace{1em}\\
  \includegraphics[width=\columnwidth]{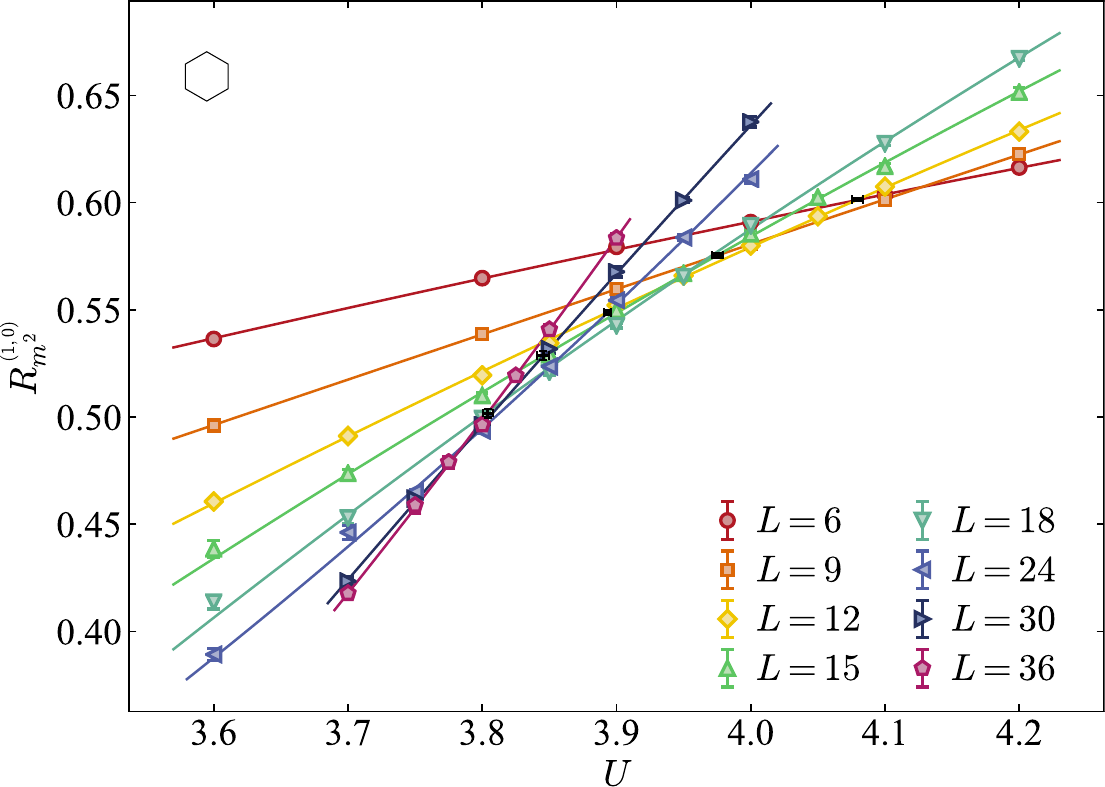}
  \caption{Correlation ratios and their crossing points (black) as obtained from the intersections of polynomial fits to pairs $(L,2L+1)$ and $(L,2L)$ of the finite-size data, for the SLAC formulation and the honeycomb, respectively.}
  \label{Xing}
\end{figure}

Here we compare the CRs $R_{m^2}^{_{(1,0)}}$, $R_{m^2}^{_{(1,1)}}$, and $R_{m^2}^{_{(2,0)}}$ from simulations of the Hubbard model at half filling with the SLAC formulation and on the honeycomb lattice, with the focus on $R_{m^2}^{_{(1,0)}}$ and $R_{m^2}^{_{(1,1)}}$, which are dominated by long wave length correlations. In \cfig{Xing}(a),(b) we present the data for $R_{m^2}^{_{(1,0)}}$ for SLAC fermions and the honeycomb (for additional CRs see \capp{app:raw}). 

\begin{figure}[t]
  \centering
  \includegraphics[width=0.97\columnwidth]{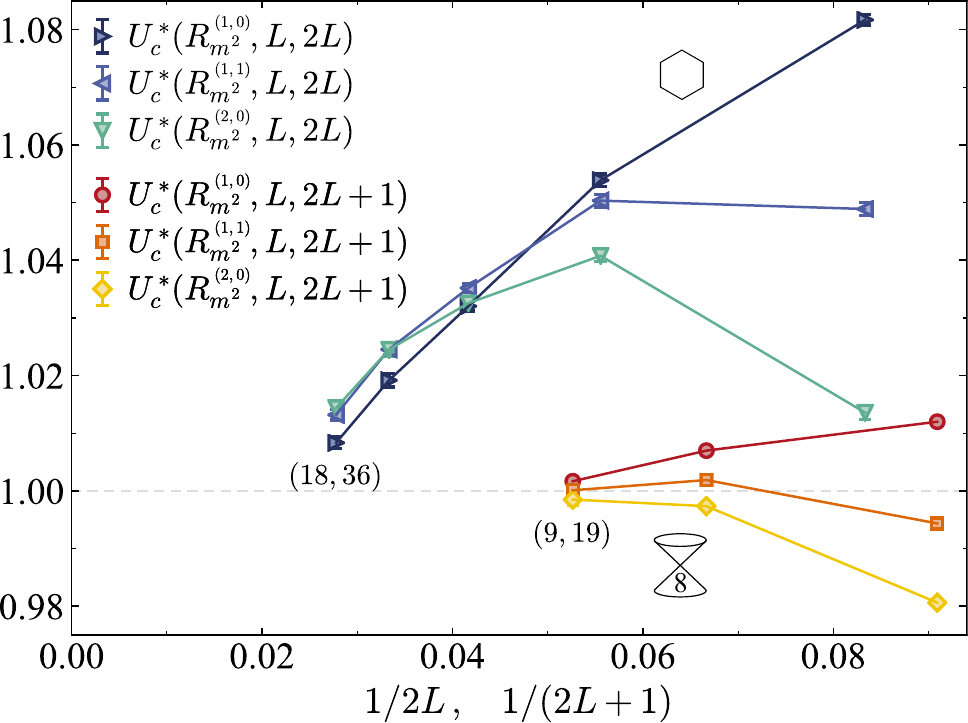}
  \caption{Three correlation ratios from the SLAC formulation and the honeycomb lattice close to the critical point for different system sizes normalized to approach unity. Results from SLAC fermion simulations converge significantly faster.}
  \label{Xing_comparison}
\end{figure}

We first determine the critical point from the finite-size intersections of pairs of CRs. We interpolate the data by second order polynomials (although predominantly linear) for each system size and then determine their intersections and the associated variance from a bootstrapping analysis. For the SLAC formulation we track the crossing points ${R_{m^2}(u,L) = R_{m^2}(u,2L+1)}$, for the honeycomb we track ${R_{m^2}(u,L) = R_{m^2}(u,2L)}$. We normalize their values with respect to the estimates of the critical points ${U_c^* = U_c/5.98}$ for the SLAC formulation and ${U_c^* = U_c/3.77}$ for the honeycomb lattice (c.f. \csec{FSS2}), such that both should approach unity as ${L^{-1/\nu-\min(\omega,2-z-\eta_{\phi})}}$ for sufficiently large lattices \cite{Campostrini14}. While all dimensionless ratios should follow the same universal scaling behavior their approach can differ by nonuniversal factors. The comparison in \cfig{Xing_comparison} shows that the crossing points of the SLAC formulation significantly converge faster to the critical point despite the smaller lattice sizes. The figure further illustrates that all choices of the CRs approach the same critical point for large lattices.

\begin{figure}[t]
  \centering
  \includegraphics[width=\columnwidth]{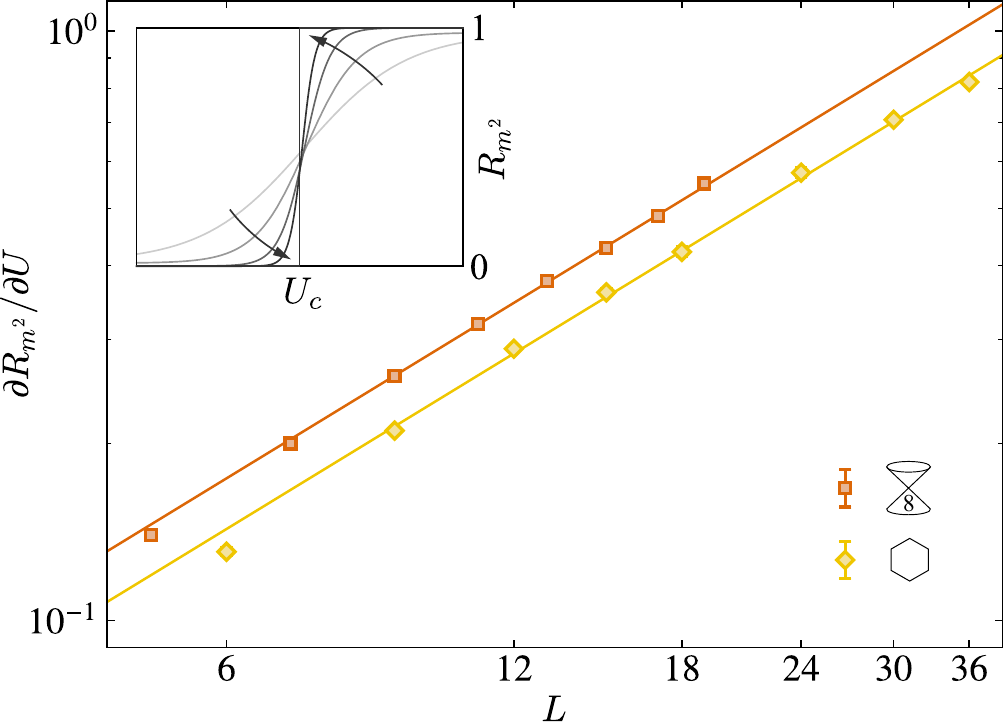}
  \caption{The slope of the CRs close to the critical point for different system sizes indicates how rapidly the ratios approach the the discontinuity at $U_c$ in the TDL, sketched in the inset.}
  \label{Xing_slopes}
\end{figure}

Provided one has data close enough to the critical point, such that the scaling function can be approximated by a linear function, the slope of the CR ${\partial R_{m^2}/\partial u \sim a_1 L^{1/\nu}}$ close to the critical point may be used as a measure of how fast the CR approaches the critical point and jumps from 0 to 1 in the TDL. The coefficient $a_1$ in \ceqn{eq:ansatz} then determines how fast their slope diverges as $R_{m^2}$ becomes discontinuous as a function of the system size. Figure~\ref{Xing_slopes} illustrates the advantage of a larger $a_1$ in the SLAC formulation, i.e., the faster convergence with system size. The lines are linear fits to the finite-size slopes on a double logarithmic scale and reflect the approximate system size dependence ${\propto L^{1/\nu}}$ \footnote{The difference of slopes at this scale is indistinguishable for different choices of $R_{m^2}^{_{(n_1,n_2)}}$ within the same lattice type.}. This also provides a rough estimate for the critical exponent ${\nu\approx 1}$ which is shared among the two lattices.

The two measures, the slope of the CRs close to criticality, which dominates the first term in \ceqn{eq:ansatz}, and the drift of the crossing points captured by scaling corrections in the second term show the SLAC formulation leads to more favorable finite-size behavior.

\subsection{finite-size scaling analysis\label{FSS2}}

We now apply the finite-size scaling (FSS) ansatz for the finite-size data of $R_{m^2}(u,L)$
\begin{eqnarray}
   R_{m^2}(u,L) &=& R_{m^2}^* + f_0^R(w) + L^{-\omega} f_1^R(w)\nn\\
   &=& R_{m^2}^* + \sum_{n=1}^{n_\text{max}}a_n w^n + L^{-\omega} \sum_{m=0}^{m_\text{max}}b_m w^m\;,
   \label{eq:ansatz}
\end{eqnarray}
to fit the CR data in \cfig{Xing} in the range ${U\in [5.79,6.23]}$ in order to extract the critical exponent of the correlation length $\nu$ and the critical coupling $U_c$. Here, ${w = u L^{1/\nu}}$, and we series expand the scaling functions $f_0^R$ and $f_1^R$ in polynomials \cite{Campostrini14}. The exponent $\omega$ accounts for irrelevant scaling contributions and the drift of the crossing points. For sufficiently large lattices contributions from the second term become negligible. A single nonlinear curve fit to the data has the tendency to significantly underestimate the statistical error of the fit parameters. In addition, the convergence of the parameters to the globally optimal fit is not guaranteed. Instead, we perform bootstrap sampling of the raw data, i.e., we carry out $10^5$ fits, where we vary the data randomly chosen from a normal distribution centered around their mean values with a squared variance corresponding to their errorbars and random initial fit parameters in a wide range, which is significantly larger than the final distribution of the resulting fit parameters. The variance of the bootstrap samples is then used to calculate the statistical errors.

\begin{figure}[t]
  \centering
  \includegraphics[width=1.\columnwidth]{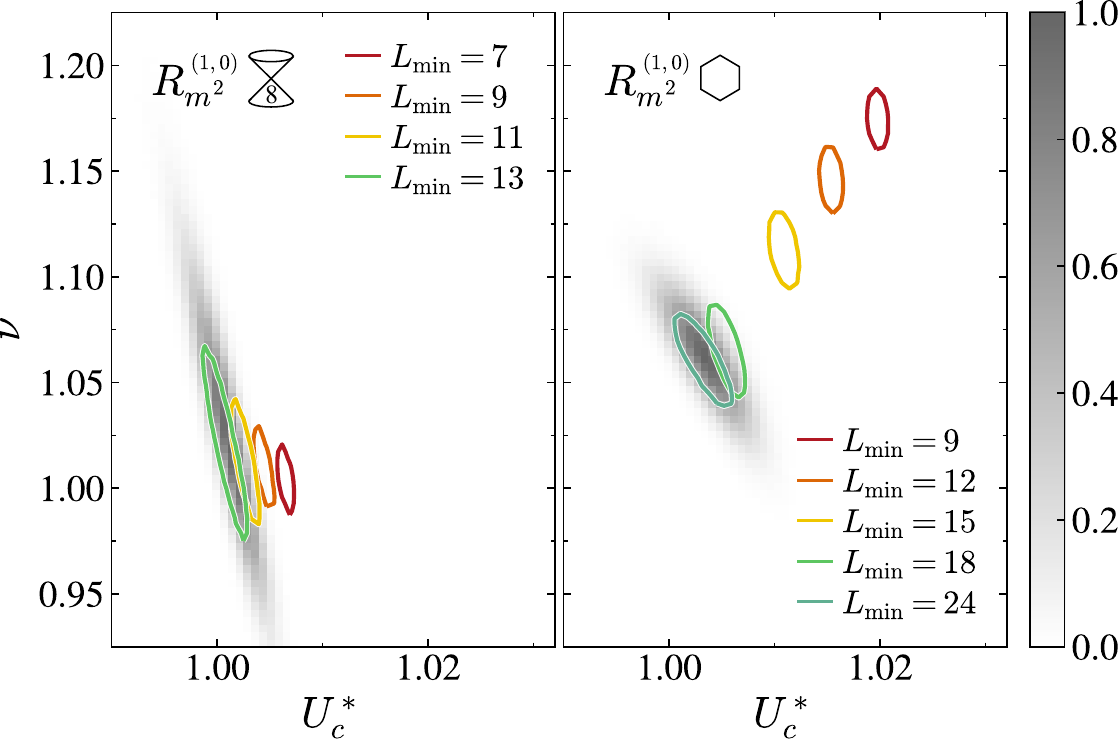}
  \caption{Histograms of the fit results to the correlation ratio close to the critical point from which we extract the critical exponent $\nu$ and $U_c$ from fits to the data. Ellipses indicate the standard deviation of the distributions for varying $L_{\rm min}$.}
  \label{fig:nuUc}
\end{figure}

We consider the scaling ansatz without corrections to scaling first, i.e., ${f_1^R = 0}$. Including only system sizes ${L\ge L_\text{min}}$ in the analysis we obtain the distribution of the fit solutions in \cfig{fig:nuUc} with a single basin of solutions for fit parameters $\nu$ and $U_c$ (Fits from additional CRs are provided in \capp{app:raw}). The squared variances of the distributions for different choices of $L_\text{min}$ are indicated by the ellipses. Independent of the lattice type a correlation of the fit parameters is visible in the skewed distribution, such that smaller values of $\nu$ are more likely to be associated with larger values of $U_c$ and vice versa. We observe stable fits for the critical coupling and critical exponent $\nu$ across the different parameter sets. The fits do not significantly change for expansion orders $n_\text{max}\ge 1$, which supports that the considered coupling range is close enough to the critical point to observe the dominant linear behavior of the CRs. If we allow for scaling corrections from irrelevant terms ${f_1^R \ne 0}$ we actually see that the additional degrees of freedom appear negligible for the SLAC fermions and tend to destabilize the fit as there is little drift in the finite-size crossing points (c.f. \cfig{Xing_comparison} and \capp{app:raw}). The fits without scaling corrections allow for a direct comparison of convergence between the two lattice types. Where the SLAC fermions show little drift in the critical coupling $U_c$ and the exponent $\nu$, the estimates from the honeycomb lattice clearly move reflecting the crossing point analysis in \cfig{Xing_comparison}. For large lattice sizes the estimates for $\nu$ coincide within error bars.

\begin{figure}[t]
  \centering
  \includegraphics[width=\columnwidth]{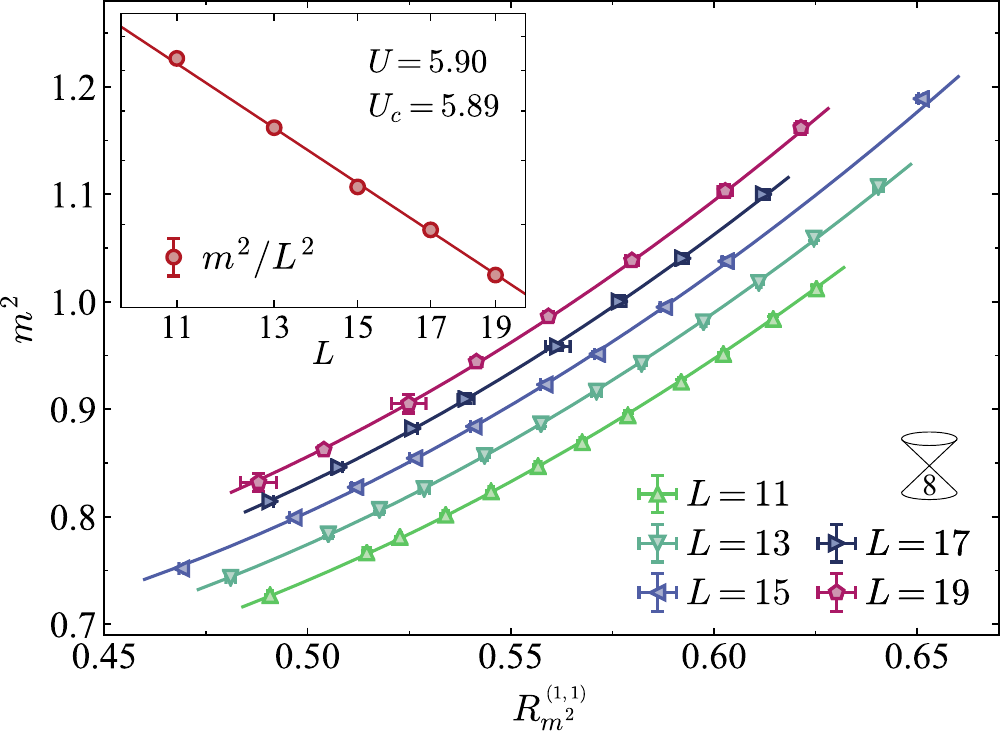}\vspace{1em}\\
  \includegraphics[width=\columnwidth]{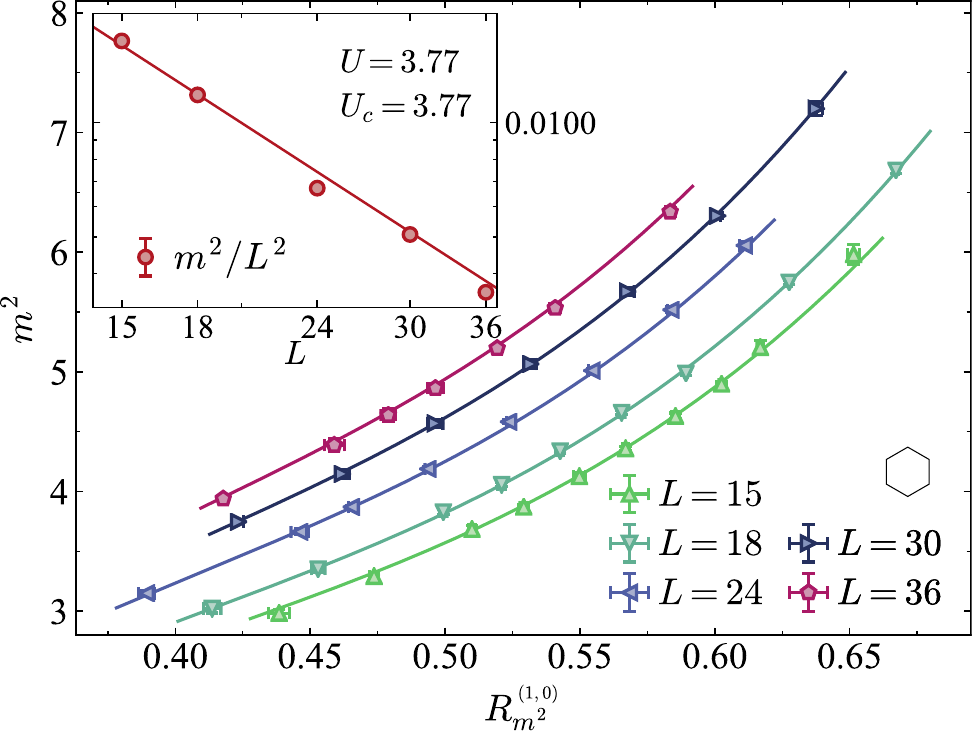}
  \caption{Fit of the finite-size scaling ansatz to the squared magnetization in order to extract the bosonic anomalous dimensions $\eta_{\phi}$ for the SLAC formulation (top) and the honeycomb lattice (bottom). The insets illustrate the compatibility of the estimated exponents with the finite-size decay behavior of the correlations at the largest distance.}
  \label{eta_phi}
\end{figure}

\begin{figure}[t]
  \centering
  \includegraphics[width=0.92\columnwidth]{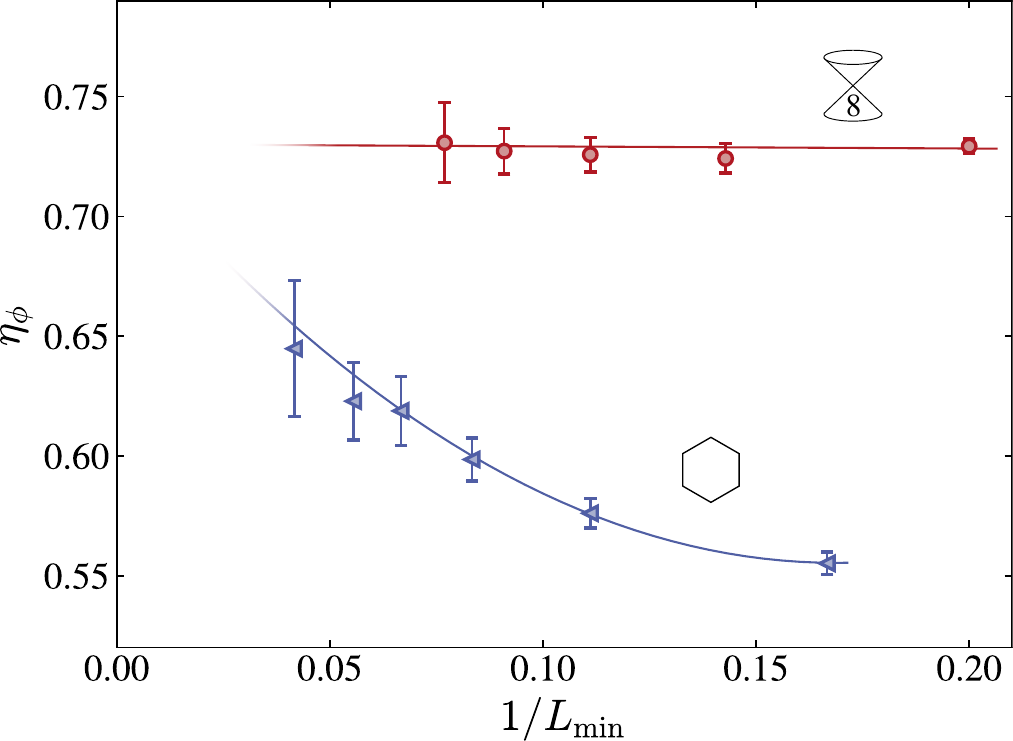}
  \caption{The boson anomalous dimension $\eta_{\phi}$ from fits with the scaling ansatz to the SLAC and honeycomb data for $R_{m^2}^{_{(1,0)}}$ as in \cfig{eta_phi}, while considering only lattice sizes ${L \ge L_\text{min}}$. Lines are guide to the eye only. 
  \label{eta_phiLmin}}
\end{figure}

The critical exponent of the correlation function associated with the order parameter $\eta_{\phi}$ can be extracted from the scaling the squared magnetization as a function of the CR $R_{m^2}$. This proves to be more stable than from the slope of $m^2$ as a function of system size, due to the larger amount of data to fit without the need for precise estimates of the critical point, thus eliminating the exponent $\nu$ from scaling. The FSS ansatz reduces to
\begin{align}
	m^2(R_{m^2},L) = L^{-(1+\eta_{\phi})} f_0^m(R_{m^2})\;,
\end{align}
where we assume ${z=1}$ \cite{Campostrini14,Toldin15}. Fits to the data are shown in \cfig{eta_phi}, where the insets illustrate the decay ${m^2 \sim L^{-(1+\eta_{\phi})}}$ using the extracted $\eta_{\phi}$ from the main panels. 
As we track the evolution of the fits results with increasing minimum system size ${L \ge L_\text{min}}$ in \cfig{eta_phiLmin}, we see that $\eta_{\phi}$ approaches the same value for both lattices. Notably, the estimate from the SLAC formulation converges significantly faster, again suggestion improved finite-size behavior of the SLAC simulations.

We repeat the procedure to extract the anomalous dimension of the fermions $\eta_{\Psi}$ from the off-diagonal elements of the single-particle Green's function at the smallest lattice momentum ${G_{ab}(\delta\mathbf{k})=\langle a^{\dagger}_{\delta\mathbf{k}} b^{\phantom{\dagger}}_{\delta\mathbf{k}}\rangle}$
with the FSS ansatz
\begin{align}
	G_{ab}(\delta\mathbf{k})(R_{m^2},L) = L^{-\eta_{\Psi}} f_0^G(R_{m^2})\;,
\end{align}
again under the assumption of ${z=1}$. 

The extracted estimates for the critical exponents from the SLAC formulation read ${\nu = 1.02(3)}$ for the correlation function, ${\eta_{\phi}=0.73(1)}$ for the anomalous dimension of the correlations associated with the order parameter and  ${\eta_{\psi}=0.09(1)}$ for the anomalous dimension of the fermion correlations. The results coincide for the two lattice types once sufficiently large honeycomb lattices are considered. The consistency of the exponents between the two lattice types using the same scaling analysis favors the SLAC formulation for the superior finite-size behavior. Note, the chiral Heisenberg exponents extracted here are incompatible with the universality of a long-range model O(3) spin models \cite{Defenu17,Song24}.

\section{Conclusion\label{conclusion}} 

In this manuscript we have discussed the implementation and details of SLAC fermions in the Hamiltonian formulation, their benefits and drawbacks in simulations. The SLAC fermions represent an effective low energy model and before their application, their operative range and restrictions should be carefully considered. Here, these are mostly the dimensionality of the considered system and the type of the interactions involved. We have investigated the influence of dynamically induced long-range super-exchange on the phases as well as the critical point and provided arguments for a long-range ordered phase, which breaks SU(2) symmetry, at finite temperature without violating the MWH theorem. The validated accurate description at strong coupling by means of linear spin wave theory allowed us to identify gapless Goldstone modes in accordance with spontaneous symmetry breaking of a continuous symmetry. 

The system at the critical point is at least non-perturbatively renormalizable and the extracted critical exponents presented here and now in several cases of simulations with a single Dirac cone in 2+1$d$ appear to be in accordance with existing estimates from local models, in particular in cases where we have exact results to compare with \cite{Li18,Zerf17,Lang19,Tabatabaei22,Erramilli23,Xu24}. Most importantly, they  significantly reduce finite-size effects, which are a major limiting factor in QMC simulations. Moreover, the SLAC formulation may be modified for a variety of low energy dispersions, which might be realized in local models, but where the physics in interacting systems is hidden at small energy scales that cannot be resolved in accessible lattice sizes of local Hamiltonians.

Pathological effects of SLAC fermions in 2+1$d$ at ${T=0}$ cannot not be ruled out completely and more investigations are certainly called for. Yet, neither the anomalous Goldstone modes, nor the finite temperature AFM order imply that the nature of the QCP, i.e., its critical properties, differ from the chiral Heisenberg GNY universality class or the QCP found in local models. If one considers the anisotropic AFM a different type of fixed point than the AFM on the honeycomb, albeit with the same remaining symmetries, they can still share the same critical properties: The working assumption of GNY universality remains that critical properties depend on the dimensionality, the symmetry of the order parameter and number of fermions. Furthermore, long-range hopping in the Hamiltonian does not exclude a conformal fixed point. On the contrary, effective actions obtained from integrating out, e.g., gauge fields, often result in non-local, yet conformal field theories.

As such, the use of SLAC fermions in 2+1$d$ remains an intriguing, non-trivial issue with strong positive evidence in favor of the SLAC formulation. We attribute this to the optimal kinetic term, which produces a perfect Dirac cone across the whole BZ, rather than only for a limited portion of available momenta and as such minimizes non-relativistic contributions from the noninteracting dispersion.

Future investigations include the effective spin model which emerges at strong coupling as worst case test of the influence of the major axis long-range coupling on the critical exponents, the comparison of the critical spectrum and scaling dimensions as fingerprint independent of critical exponents \cite{Schuler21}, and further comparison with local approximations of Dirac fermions which attempt to optimize the extent of the relativistic dispersion without fermion doubling such as Wilson- and tangent-fermions \cite{Wilson74,Beenakker23}.

\begin{acknowledgments}
We thank F.~F. Assaad for valuable discussions. We gratefully acknowledge the allocation of CPU time on the HPC infrastructure LEO of the University of Innsbruck and the Vienna Scientific Cluster VSC for the allocation of CPU time.
\end{acknowledgments}
\vspace{1em}
\begin{appendix}

\section{Hopping amplitude of SLAC Fermions\label{ap:hoppingSLAC}}

Here we calculate the discrete Fourier transformation of the hopping matrix elements for SLAC fermions on lattices of even and odd linear dimension $L$ which include the Dirac point ${k=0}$. Let us compute real-space representation of the linear dispersion ${\varepsilon(k) = k}$ in one dimension
\begin{align}
	\varepsilon(x) &= \frac{1}{L} \sum_k k \,\E^{\I k x}\;,
\end{align}
where ${k = 2\pi m/L}$ and ${m = -M, \ldots, M\in\mathbb{N}}$ with ${M=(L-1)/2}$ for odd $L$ system sizes. 
In the following we will use the exponential sum formula to replace the sum with a fraction
\begin{align}
   \sum_{k=0}^{N-1} k\,\E^{\I k x} 
   &= \frac{1}{\I}\frac{\mathrm{d}}{\mathrm{d}x} \sum_{k=0}^{N-1}  \,\E^{\I k x} \nonumber\\
   &= \frac{1}{\I}\frac{\mathrm{d}}{\mathrm{d}x} \frac{1 - \E^{\I N x}}{1 - \E^{\I x}} \nonumber\\
   &= -\frac{N\,\E^{\I N x} \left(1 - \E^{\I x}\right) - \E^{\I x}\left( 1 - \E^{\I N x} \right)}{ \left(1 - \E^{\I x}\right)^2}~,\label{eq:Apexpsum}
\end{align}
so we restrict the sum over positive momenta with ${m = 0,\ldots, M}$, such that
\begin{align}
	\varepsilon(x) &= \frac{1}{L} \sum_{k\ge 0} k \,\left( \,\E^{\I k x} - \,\E^{-\I k x} \right)
    \nonumber\\
    &= \frac{2\I}{L} \sum_{k\ge 0} k \sin\left(k x\right)\nonumber\\
    &= \frac{2\I}{L}\;\mathrm{Im} \sum_{k\ge 0} k \,\E^{\I k x}\nonumber\\
    &= \frac{4\I\pi}{L^2}\;\mathrm{Im} \sum_{m=0}^M m \,\E^{\I\frac{2 \pi m}{L}x}\;. \label{eq:Apfx}
\end{align}
We use \ceqn{eq:Apexpsum} to rewrite \ceqn{eq:Apfx} as
\begin{widetext}
\begin{align} 
	\varepsilon(x) 
	&= -\frac{4\I\pi}{L^2} \;\mathrm{Im}\left[\frac{\frac{L+1}{2}\,\E^{\I\frac{2\pi}{L} \frac{L+1}{2} x} \left(1 - \E^{\I\frac{2\pi}{L} x}\right) - \E^{\I\frac{2\pi }{L} x}\left( 1 - \E^{\I\frac{2\pi }{L}\frac{L+1}{2} x} \right)}{ \left(1 -\E^{\I\frac{2\pi }{L}x}\right)^2} \right] \nonumber \\ 
    &= -\frac{4\I\pi}{L^2} \;\mathrm{Im}\left[\frac{\frac{L+1}{2} \E^{\I\pi x} \,\E^{\I\frac{2\pi }{L}x} \left(\E^{ -\I\frac{\pi}{L}x} - \E^{\I\frac{\pi}{L}x}\right) - \E^{\I\frac{2\pi }{L}x}\left( 1 - \E^{\I\pi x} \,\E^{\I\frac{\pi}{L}x} \right)}{\,\E^{\I\frac{2\pi }{L}x} \left(\E^{ -\I\frac{\pi}{L}x} -\E^{\I\frac{\pi}{L}x}\right)^2} \right] \nonumber \\ 
    &= -\frac{4\I\pi}{L^2} \;\mathrm{Im}\left[\frac{\frac{L+1}{2} \E^{\I\pi x} (-2\I)\,\sin\left(\frac{\pi}{L}x\right) - \left( 1 - \E^{\I\pi x} \,\E^{\I\frac{\pi}{L}x} \right)}{-4\,\sin\left(\frac{\pi}{L}x\right)^2} \right] \nonumber \\ 
    &= -\frac{4\I\pi}{L^2} (-1)^x \frac{\frac{L+1}{2}\,(-2)\,\sin\left(\frac{\pi}{L}x\right) - \sin\left(\frac{\pi}{L}x\right)}{-4\,\sin\left(\frac{\pi}{L}x\right)^2} \nonumber \\ 
    &= -\frac{\I\pi}{L} \frac{(-1)^x}{\sin\left(\frac{\pi}{L}x\right)}\;. \label{ap:fx}
\end{align}
\end{widetext}
For even $L$ system sizes we have the momenta ${k = 2\pi m/L}$, where ${m = -M, \ldots, M-1\in\mathbb{N}}$ with ${M=L/2}$ and we can again use \ceqn{eq:Apexpsum} to proceed with ${N=L/2+1}$ in the same way as for \ceqn{ap:fx} to first compute
\begin{widetext}
\begin{align} 
	\varepsilon'(x) 
	&= -\frac{4\I\pi}{L^2} \;\mathrm{Im}\left[\frac{\left(\frac{L}{2}+1\right)\,\E^{\I\frac{2\pi}{L} \left(\frac{L}{2}+1\right) x} \left(1 - \E^{\I\frac{2\pi}{L} x}\right) - \E^{\I\frac{2\pi }{L} x}\left( 1 - \E^{\I\frac{2\pi }{L}\left(\frac{L}{2}+1\right) x} \right)}{ \left(1 -\E^{\I\frac{2\pi }{L}x}\right)^2} \right] \nonumber \\ 
	&= -\frac{4\I\pi}{L^2} \;\mathrm{Im}\left[\frac{\left(\frac{L}{2}+1\right)\,\E^{\I\pi x}\,\E^{\I\frac{2\pi}{L} x}\,\E^{\I\frac{\pi}{L} x} \left(\E^{-\I\frac{\pi}{L} x} - \E^{\I\frac{\pi}{L} x}\right) - \E^{\I\frac{2\pi }{L} x}\left( 1 - \E^{\I\pi x}\,\E^{\I\frac{2\pi }{L} x} \right)}{ \E^{\I\frac{2\pi }{L}x} \left(\E^{-\I\frac{\pi }{L}x} -\E^{\I\frac{\pi }{L}x}\right)^2} \right] \nonumber \\ 
	&= -\frac{4\I\pi}{L^2} \;\mathrm{Im}\left[\frac{\left(\frac{L}{2}+1\right)\,\E^{\I\pi x}\,\E^{\I\frac{\pi}{L} x} (-2\I) \,\sin\left(\frac{\pi}{L} x\right) - \left( 1 - \E^{\I\pi x}\,\E^{\I\frac{2\pi }{L} x} \right)}{-4\,\sin\left(\frac{\pi}{L} x\right)^2} \right] \nonumber \\ 
	&= -\frac{4\I\pi}{L^2} (-1)^x \frac{\left(\frac{L}{2}+1\right)\,(-2)\,\cos\left(\frac{\pi}{L} x\right) \,\sin\left(\frac{\pi}{L} x\right) + 2\sin\left(\frac{\pi}{L} x\right)\cos\left(\frac{\pi}{L} x\right)}{-4\,\sin\left(\frac{\pi}{L} x\right)^2} \nonumber \\ 
    &= -\frac{\I\pi}{L} (-1)^x \frac{\cos\left(\frac{\pi}{L}x\right)}{\sin\left(\frac{\pi}{L}x\right)}\;,\end{align}
\end{widetext}
and then subtract from $\varepsilon'(x)$ the contribution ${m=L/2}$ to avoid double counting
\begin{align}
	\varepsilon(x) &= \frac{4\I\pi}{L^2}\;\mathrm{Im} \left[\sum_{m=0}^M m \,\E^{\I\frac{2 \pi m}{L}x} \right] - \frac{\pi}{L}\,\E^{\I\pi x} \nonumber\\
	&= -\frac{\I\pi}{L} (-1)^x \frac{\cos\left(\frac{\pi}{L}x\right)}{\sin\left(\frac{\pi}{L}x\right)} - \frac{\pi}{L}\,(-1)^x \nonumber\\
    &= -\frac{\I\pi}{L} (-1)^x \left(\frac{\cos\left(\frac{\pi}{L}x\right)}{\sin\left(\frac{\pi}{L}x\right)} + \I\right)\;.
\end{align}
For the linearity of the Fourier transformation the same calculation holds for the two-dimensional case and  the hopping amplitude for distance ${\mathbf{r} = (x, y)}$ is given by
\begin{align}
t(\mathbf{r}) &= \frac{1}{L^2}\sum_{\mathbf{k}} \left( k_x + \I k_y \right) \,\E^{\I \mathbf{k}\cdot\mathbf{r}}
\nonumber \\
&= \frac{1}{L^2} \left( \sum_{\mathbf{k}} k_x \,\E^{\I \mathbf{k}\cdot\mathbf{r}} + \I \sum_{\mathbf{k}} k_y \,\E^{\I \mathbf{k}\cdot\mathbf{r}} \right)
\nonumber \\ 
&= \varepsilon(x)\,\delta_{y,0} + \I\,\varepsilon(y)\,\delta_{x,0}\;.
\end{align}

\section{Additional correlations ratios \& finite-size scaling\label{app:raw}}

\begin{figure*}[t]
  \centering
  \includegraphics[width=\textwidth]{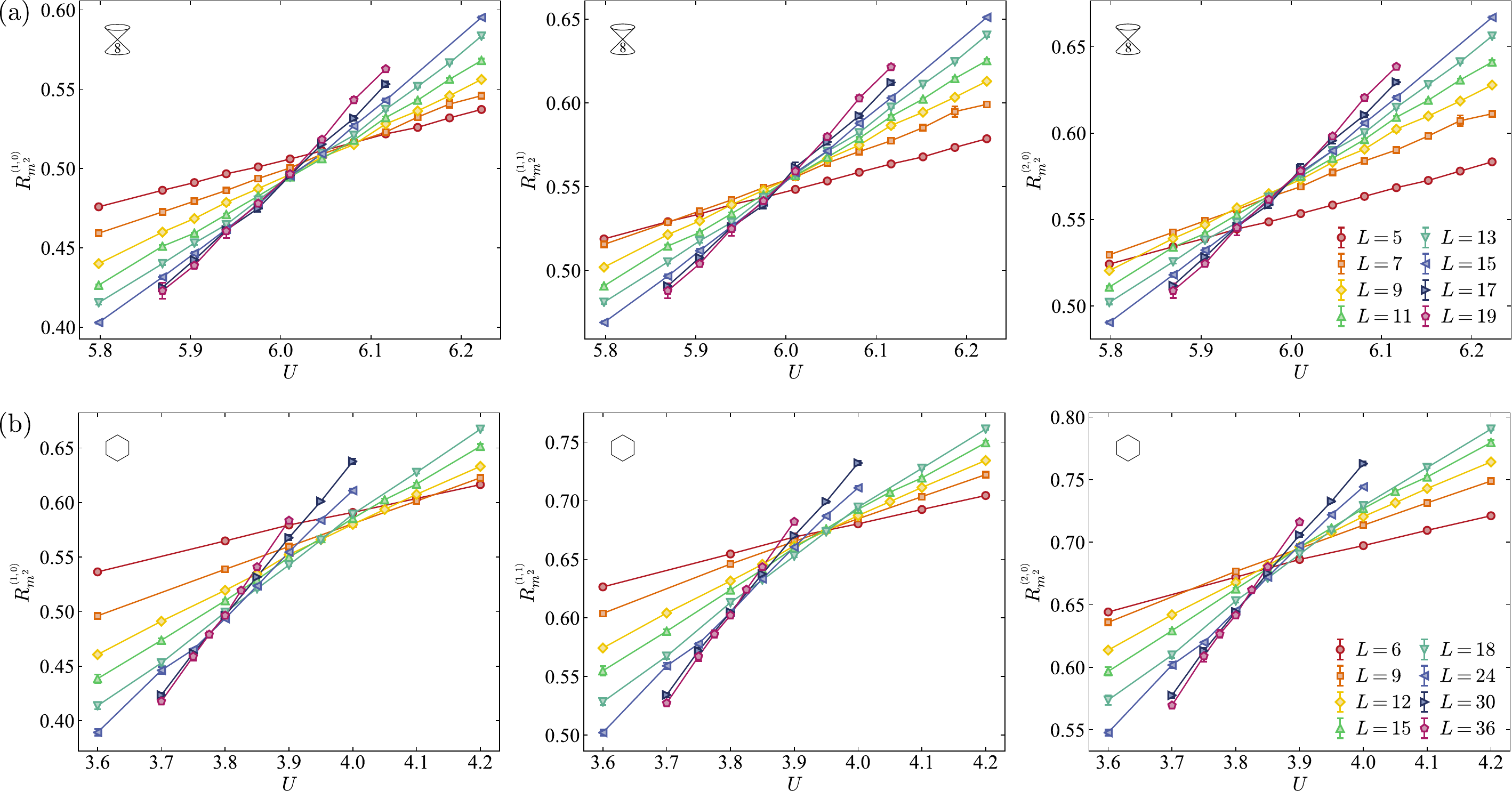}\\
  \caption{The correlation ratios for the moment $(\mathbf{b}_1,0)$, $(\mathbf{b}_1,\mathbf{b}_2)$ and $(2\mathbf{b}_1,0)$ away from the Dirac nodal point for (a) the ${N=8}$ SLAC formulation and (b) the honeycomb lattice. }
  \label{app:Rm2dk_raw}
\end{figure*}

\begin{figure}[t]
  \centering
  \includegraphics[width=1.\columnwidth]{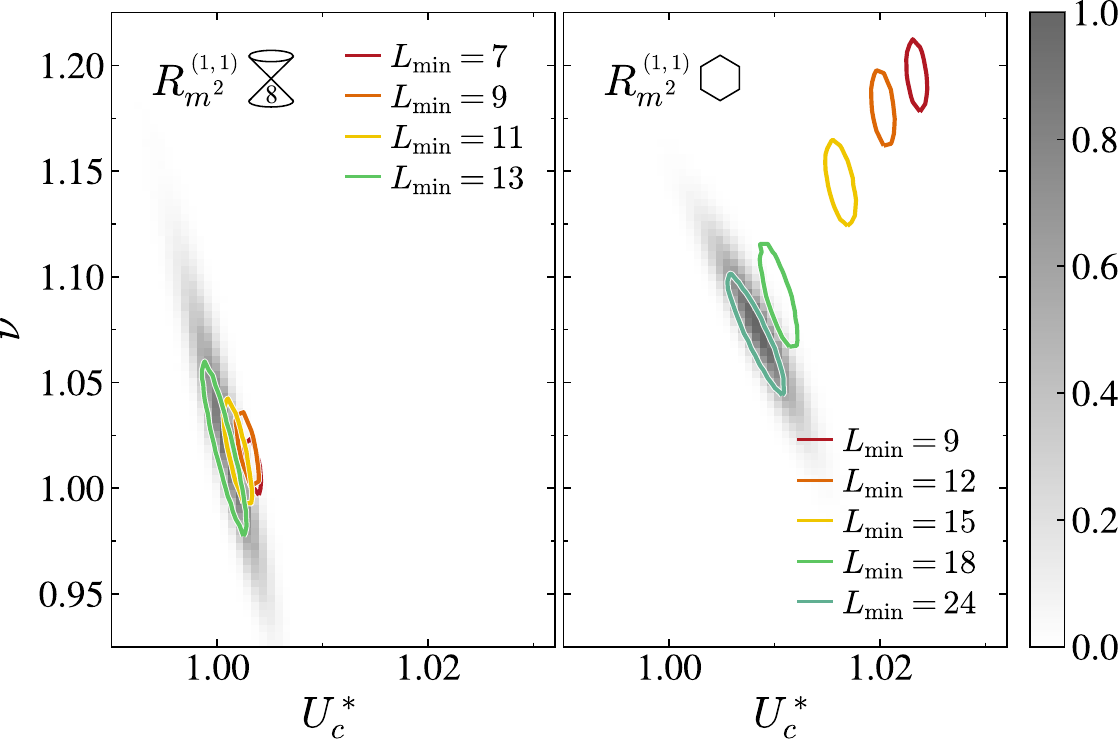}\\
  \includegraphics[width=1.\columnwidth]{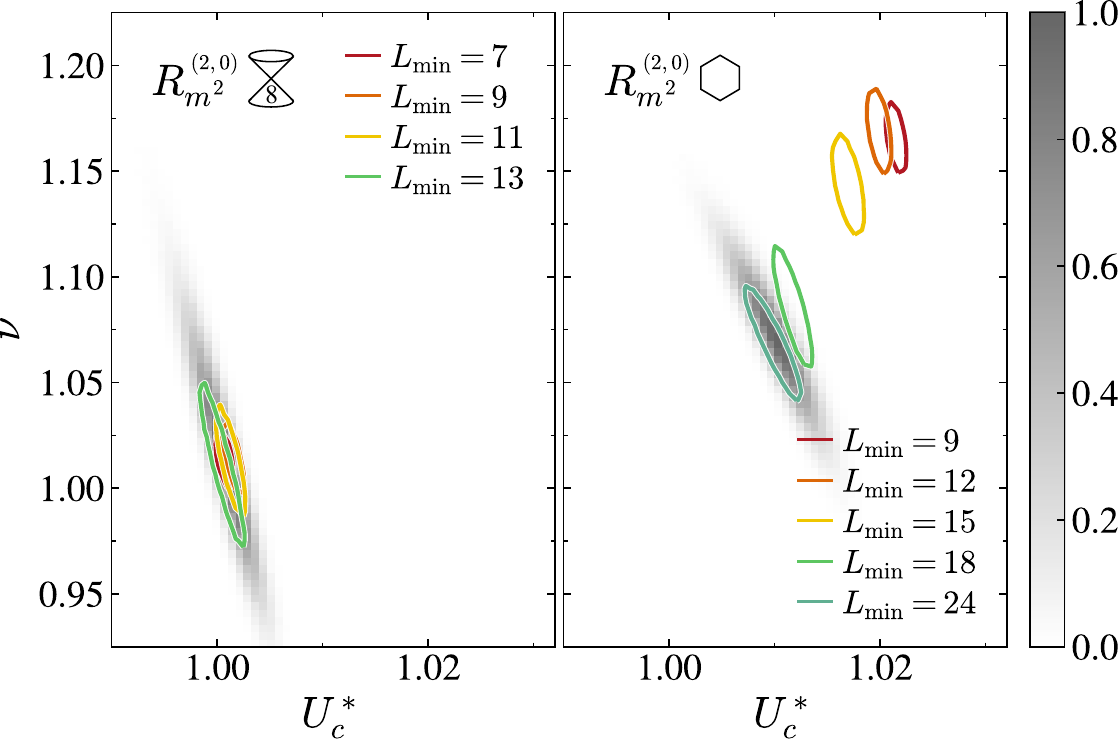}
  \caption{Histograms of the fit results to the correlation ratio close to the critical point from which we extract the critical exponent $\nu$ and $U_c$ from fits to the data. Ellipses indicate the standard deviation of the distributions for varying $L_{\rm min}$.}
  \label{app:nuUc}
\end{figure}

The CRs for the momenta $(\mathbf{b}_1,0)$, $(\mathbf{b}_1,\mathbf{b}_2)$ and $(2\mathbf{b}_1,0)$ close to the Dirac nodal point for the ${N=8}$ SLAC formulation and the honeycomb lattice are presented without fits in \cfig{app:Rm2dk_raw}. Lines are guides to the eye only. Fit results to the CRs without scaling corrections on both lattices are shown in \cfig{app:nuUc}.

\begin{figure}[t]
  \centering
  \includegraphics[width=\columnwidth]{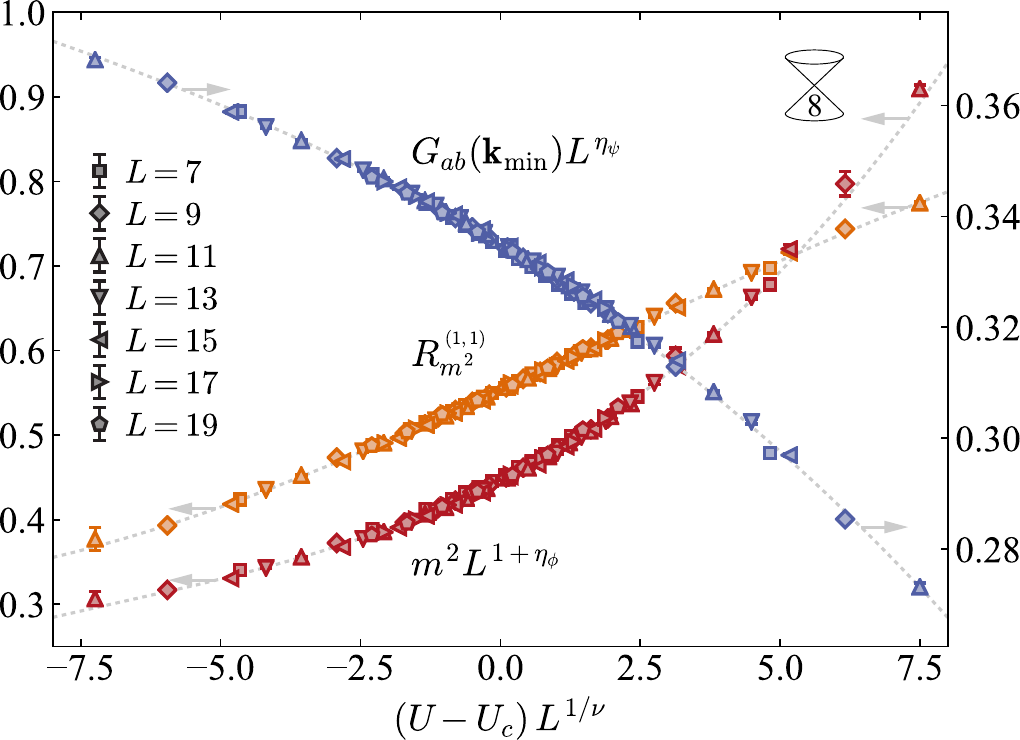}\\
  \caption{Data collapse of the correlation ratio $R_{m^{2}}^{(1,1)}$, the squared magnetization $m^2$ (left scale) and the single particle Green's function $G_{ab}(\delta\mathbf{k})$ at the smallest momentum (right scale) from the SLAC formulation.}
  \label{app:collapse}
\end{figure}

The estimates ${\nu = 1.02}$, ${\eta_{\phi}=0.73(1)}$ and ${\eta_{\psi}=0.09(1)}$ are used to collapse the SLAC data of the correlation ratio $R_{m^{2}}^{(1,1)}$, the squared magnetization $m^2$ and the single particle Green's function $G_{ab}(\delta\mathbf{k})$ at the smallest momentum in \cfig{app:collapse}.

\end{appendix}

\bibliography{paper.bib}

\end{document}